\documentclass{jpsj2}

\def\runauthor{Gen'ya \textsc{Sakurai} and Yoshio \textsc{Kuramoto}}

\title{%
Multipolar Interactions in the Anderson Lattice with Orbital Degeneracy}

\author{%
Gen'ya \textsc{Sakurai}$^{1}$\thanks{E-mail address: genya@cmpt.phys.tohoku.ac.jp} and Yoshio \textsc{Kuramoto}}

\inst{%
Department of Physics, Tohoku University, Sendai 980-8578}

\recdate{\today}

\abst{%
Microscopic investigation is performed for intersite multipolar interactions in the orbitally degenerate Anderson lattice, with CeB$_6$ taken as an exemplary target.
In addition to the $f^0$ intermediate state, 
$f^2$ Hund's-rule ground states are included as intermediate states for the interactions. 
The conduction-band states
are taken as plane waves and the hybridization as spherically symmetric.
The spatial dependences of multipolar interactions are given by
the relative weight of partial wave components along the pair of sites.
It is clarified how the the anisotropy arises in the interactions depending on the orbital degeneracy and the spatial configuration.
The stability of the $\Gamma_5$ antiferro-quadrupole order in the phase II of CeB$_6$ is consistent with our model.
Moreover, the pseudo-dipole interactions follow a tendency required by the phenomenological model for the phase III. 

}

\kword{quadrupole order, orbital degeneracy, pseudo-dipole, octupole,
Anderson lattice, RKKY interaction
}

\begin{document}
\sloppy
\maketitle
\section{Introduction}

Rare-earth compounds with orbital degrees of freedom have been attracting much interest.
Among them, 
 CeB$_6$ shows antiferro-quadrupolar (AFQ) ordering below $T_Q$ = 3.2K called the phase II, and antiferro (AF) magnetic ordering below $T_N$ = 2.3K called the phase III.\cite{Effantin,Erkelens}
This AFQ ordering has been identified as of quadrupoles of $\Gamma_5$ type.\cite{Sakai}
On the other hand, the phase III has a complicated magnetic structure with double $\mib{k}$, i.e., $\mib{k_1} = (1/4, 1/4, 1/2)$ and $\mib{k_1}^\ast = (1/4, -1/4, 1/2)$ in units of $2\pi/a$ with $a$ being lattice constant, as has been probed by neutron scattering.\cite{Effantin}
The magnetic structure of the phase III is such that parallel and anti-parallel magnetic moments of next-nearest-neighbor pairs are equally populated.
Thus the magnetic structure has no gain of energy from the isotropic exchange interaction acting on nearest neighbors.
In order to explain the stability of the structure, pseudo-dipole-type couplings have been proposed between the next-nearest-neighbors.\cite{Kusunose}  
However, microscopic justification of this type of anisotropic interactions is lacking.

There are two different sources of the intersite exchange interaction: the on-site Coulomb exchange interaction mainly between 4$f$ and 5$d$ electrons in rare-earth systems, and hybridization between 4$f$ and ligand electrons.  
The Coulomb exchange always favors on-site parallel spins.
Early studies derived its consequences neglecting the orbital degeneracy, and assuming the spherically symmetric conduction states.\cite{Kasuya}
The intersite quadrupolar interaction was derived later for the model with 
the Coulomb interaction as the source.\cite{Teitelbaum}
The most elaborate contribution for the Coulomb model is made by
ref.\citen{Schmidt}, which includes the multipoles of 4$f$ electrons and the result of band structure calculation.
On the other hand, 
intersite interactions originating from hybridization have been considered in much less detail.
Theoretical literature often concerns with isotropic interactions only, which
result from neglecting the multiplet splitting in the $f^2$ configurations.\cite{Coqblin,Ohkawa}
The intersite interaction from band structure calculation was derived in ref.\citen{Takahashi}.
Recently Shiba {\it et al.} derived the effective interionic Hamiltonian
with the full use of cubic symmetry of the conduction states, 
 but only the $f^0$ intermediate state was included.\cite{Shiba}

It has been shown that the anisotropic spin exchange arises by cooperation
 of the spin-orbit coupling and the Coulomb exchange interaction.\cite{Yildirim,Kasuya}
However, the anisotropic exchange from hybridization has hardly been studied.
Motivated by this situation, we investigate in this paper the multipolar interactions induced by hybridization microscopically.  We include the spin-orbit interaction in the intermediate states to derive possible anisotropy.
Specifically we introduce the Anderson-type lattice model with orbital degeneracy, in which 
not only $f^0$ but also $f^2$ Hund's-rule ground states are considered as intermediate states.

In order to clarify the origin of the anisotropy and the difference between multipolar interactions in a simple manner, we take a model which 
has spherically symmetric hybridization
 around each Ce ion, 
and which takes the wave functions of the conduction band as plane waves.
For the ground state of each $4f^1$ configuration, we take the $\Gamma_8$
crystalline electric field (CEF) state, bearing CeB$_6$ in mind.
Then we obtain the effective interionic Hamiltonian 
by 4th order perturbation with respect to hybridization. 
The high symmetry around each Ce ion allows us to simplify the analysis significantly.

As is well known,  the $f^0$ intermediate state tends to align the spins of $f$ and conduction electrons anti-parallel.
On the other hand, the Hund's-rule correlation works in the $f^2$ intermediate states.
The lowest intermediate multiplet, referred to as $f_{\text{H}}^2$ hereafter, tends to align the $f^1$ and conduction spins parallel because the total spin $S=1$ is realized in the $f_{\text{H}}^2$ configuration.
This competition causes decrease of the spin exchange as will be investigated in detail later.
 
This paper is organized as follows.
 In \S 2 we introduce the model modifying the 
 periodic Anderson Hamiltonian with orbital degeneracy, and calculate the effective interionic Hamiltonian by the 4th order perturbation.
 In \S 3 we evaluate the multipolar coupling constants numerically, and analyze them in detail. We comment on possible relevance of the results to actual systems.
 The summary is given in the final section.
 Appendix derives analytically the range functions of the intersite interaction within our model.
\section{Effective Hamiltonian for Intersite Interactions}
Each Ce$^{3+}$ ion has one 4$f$ electron, and the spin-orbit coupling leads to the total angular momentum $j_1 = 5/2$ in the ground state of the ion.
The operator $f_{m\sigma}^\dagger(i)$ creates an 4$f$ electron at site 
$\mib{R}_i$ with the $z$-component $m$ of the orbital angular momentum $l=3$ and spin $\sigma$.
We take conduction-band wave functions as plane waves, and introduce the  creation operator
$c_{\mib{k}\sigma}^\dagger$ 
with wave number $\mib{k}$ and spin $\sigma$.
Then our model is given by
\begin{equation}
H = H_c + H_f + H_{\text{hyb}} +  H_{\text{Coulomb}},
\end{equation}
where
\begin{align}
H_c =& \sum_{\mib{k},\sigma}\epsilon_{\mib{k}}c_{\mib{k}\sigma}^\dagger c_{\mib{k}\sigma}, \\
H_f =& \sum_{i,m,\sigma}E_f f_{m\sigma}^{\dagger}(i) f_{m\sigma}(i) 
	+\frac{\lambda}{2} \sum_{i,m,m',\sigma,\sigma'}
    \! \langle i;3m|\mib{L}|i;3m'\rangle \cdot
	\mib{\sigma}_{\sigma, \sigma'}
    f_{m\sigma}^{\dagger}(i) f_{m'\sigma'}(i), \\
H_{\text{hyb}} =& \sqrt{\frac{4\pi}{N}}\sum_{i,\mib{k},m,\sigma}
		\left[V_k Y_{3m}^\ast (\Omega_{\mib{k}})e^{i\mib{k}\cdot\mib{R}_i}f_{m\sigma}^{\dagger}(i) c_{\mib{k}\sigma} + \text{H.c.}\right]. \label{Hhyb}
\end{align}
Here $\lambda$ is the spin-orbit coupling constant, 
 while $\mib{L}$ and $\mib{\sigma}$ denote the orbital angular momentum operator and Pauli matrices respectively.
The solid angle of $\mib k$ is represented by $\Omega_{\mib{k}}$.
We assume that combination of $E_f$ and $H_{\text{Coulomb}}$ is such that 
the 4$f^1$ state is stable against $4f^0$ and $4f^2$ states.

As the intermediate states with $f^2$ configuration, we consider only
the Hund's-rule ground state $f_{\text{H}}^2$ in order to emphasize the origin of the anisotropy.
In general, $f_{\text{H}}^2$ is characterized by the quantum numbers $L,S,J$ and 
$J_z=M$.
We take $L=5, S=1$ and $J=4$ according to the Hund's-rule.
Then we replace $H_f + H_{\text{Coulomb}}$ as follows:
\begin{align}
& H_f + H_{\text{Coulomb}} \nonumber \\
&\rightarrow
\sum_i E_0 |i; 0\rangle \langle i; 0|
+ \sum_{i,m_1} E_1 |i; j_1 m_1 \rangle \langle i; j_1 m_1 | 
+\sum_{i,M} E_2 |i; (LS)J M \rangle \langle i; (LS)J M |,
\end{align}
where $E_1$ is the energy of the 4$f^1$ ground state, while $E_0$ and $E_2$ are the energies of $f^0$ and $f_{\text{H}}^2$, respectively.
Experimentally, the position of the $4f^1$ level 
is about 3 to 4 eV below the Fermi surface 
in CeB$_6$ as probed by photoemission.\cite{IPES,Chi}   
On the other hand, the $4f^2$ levels have been probed by Bremsstrahlung, and they are also 3 to 4 eV above the Fermi level.\cite{IPES}  
Thus naively thinking, intermediate states $4f^0$ and $4f^2$ have comparable contribution to the effective exchange interaction.  It is not clear, however, the weight of the $4f_{\text H}^2$ relative to other contributions.  Our theory with 
only $4f_{\text H}^2$ taken into account may give the upper bound of the anisotropy in the exchange.

In eq.(\ref{Hhyb}), we then need to consider the matrix elements between  $f^0$, $f_{\text{H}}^2$ and 4$f^1$ states:
\begin{equation}
H_{\text{hyb}} \rightarrow H_{\text{hyb}}(f^0)+H_{\text{hyb}}(f^2),
\end{equation}
where $H_{\text{hyb}}(f^0)$ is the hybridization Hamiltonian between 4$f^1$ state and $f^0$, while $H_{\text{hyb}}(f_{\textrm{H}}^2)$ is that between 4$f^1$ state and $f_{\text{H}}^2$.
They are represented by
\begin{align}
H_{\text{hyb}}(f^0)
&= \sqrt{\frac{4\pi}{N}}\sum_{i,\mib{k}}\sum_{m_1,m,\sigma,\sigma_1}
	\left[
	V_k Y_{3m}^\ast(\Omega_{\mib{k}})e^{i\mib{k} \cdot\mib{R}_i} \right.
	\nonumber \\
	& \times
	\langle j_1 m_1|3\,m_1\!-\!\sigma_1,\frac{1}{2}\,\sigma_1 \rangle 
	\langle m_1\!-\!\sigma_1, \sigma_1| f_{m\sigma}^\dagger | 0\rangle
	\nonumber \\
	& \times
	\left. |i;j_1 m_1 \rangle \langle i;0| c_{\mib{k}\sigma}
	 + \text{H.c.}
	 \right], \label{Hhybf0J} \\
H_{\text{hyb}}(f_{\textrm{H}}^2)
&= \sqrt{\frac{4\pi}{N}}\sum_{i,\mib{k}}\sum_{M,m_1,m,\sigma,\sigma_1}
		\left[ V_k Y_{3m}^\ast(\Omega_{\mib{k}})e^{i\mib{k}\cdot\mib{R}_i} \right. \nonumber \\
		& \times \langle (L,S)JM|f_{m\sigma}^{\dagger}|m_1\!-\!\sigma_1,\sigma_1 \rangle
		\langle 3\,m_1\!-\!\sigma_1, \frac{1}{2}\,\sigma_1|j_1 m_1 \rangle
		\nonumber \\
		& \times
		\left. |i;(L,S)JM \rangle \langle i;j_1 m_1 | c_{\mib{k}\sigma}
		+ \text{H.c.}
		\right]. \label{Hhybf2J}
\end{align}

\subsection{Effective on-site exchange
Hamiltonian}
Using eqs.(\ref{Hhybf0J}) and (\ref{Hhybf2J}), the effective on-site exchange
Hamiltonian is obtained by the second order perturbation theory as:\cite{Hirst}
\begin{align}
H_{\text{exch}} \quad &= H_{\text{exch}}(f^0) + H_{\text{exch}}(f_{\textrm{H}}^2),\nonumber \\
H_{\text{exch}}(f^0) &= -\frac{4\pi}{N}
\sum_{i,\mib{k},\mib{k}'}
\frac{V_k V_{k'}^\ast}{E_1-E_0-\epsilon_F}
e^{i(\mib{k}-\mib{k}') \cdot \mib{R}_i}
\sum_{\text{All } m's, \sigma's}
Y_{3m_4}^\ast (\Omega_{\mib{k}}) Y_{3m_3}(\Omega_{\mib{k}'}) \nonumber \\
& \quad \times
\langle m_1\!-\!\sigma_1, \sigma_1 | f_{m_4 \sigma_4}^\dagger | 0\rangle
\langle 0| f_{m_3 \sigma_3} |m_2\!-\!\sigma_2, \sigma_2 \rangle
\nonumber \\
&\quad \times
\langle j_1 m_1|3\,m_1\!-\!\sigma_1, \frac{1}{2}\,\sigma_1 \rangle
\langle 3\,m_2\!-\!\sigma_2, \frac{1}{2}\,\sigma_2|j_1 m_2 \rangle
\nonumber \\
& \quad \times
|i; j_1 m_1 \rangle \langle i; j_1 m_2|
c_{\mib{k}'\sigma_3}^\dagger c_{\mib{k}\sigma_4}, \label{Hexchf0}\\
H_{\text{exch}}(f_{\textrm{H}}^2) &= \frac{4\pi}{N}
\sum_{i,\mib{k},\mib{k}'}
\frac{V_k V_{k'}^\ast}{E_1-E_2+\epsilon_F}
e^{i(\mib{k}-\mib{k}')\cdot \mib{R}_i}
\sum_{\text{All } m's, \sigma's}
Y_{3m_4}^\ast (\Omega_{\mib{k}}) Y_{3m_3}(\Omega_{ \mib{k}'}) \nonumber \\
& \quad \times \sum_M
\langle m_1\!-\!\sigma_1, \sigma_1|f_{m_3 \sigma_3}|(LS)JM \rangle
\langle (LS)JM|f_{m_4 \sigma_4}^\dagger |m_2\!-\!\sigma_2, \sigma_2 \rangle
\nonumber \\
&\quad \times
\langle j_1 m_1|3\,m_1\!-\!\sigma_1, \frac{1}{2}\,\sigma_1 \rangle
\langle 3\,m_2\!-\!\sigma_2, \frac{1}{2}\,\sigma_2|j_1 m_2 \rangle
\nonumber \\
& \quad \times
|i; j_1 m_1 \rangle \langle i; j_1 m_2 |
c_{\mib{k}'\sigma_3}^\dagger c_{\mib{k}\sigma_4}, \label{Hexchf2}
\end{align}
with $\epsilon_F$ being the Fermi energy.
In the above, $H_{\text{exch}}(f^0)$ and $H_{\text{exch}}(f_{\textrm{H}}^2)$ are the effective
exchange Hamiltonian with the virtual states $f^0$ and $f_{\text{H}}^2$, respectively.

Let us inspect the sign of the effective Hamiltonian (\ref{Hexchf0}) and (\ref{Hexchf2}).
If the multiplet splitting is neglected in the 4$f^2$ states, both
$f^0$ and 4$f^2$ give the antiferro-magnetic spin exchange interaction, as is well known for the Anderson model.
If the Hund's-rule splitting is included, $f_{\text{H}}^2$ favors the ferromagnetic spin exchange.
This is demonstrated as follows:
from the effective Hamiltonian (\ref{Hexchf2}), 
spin exchange corresponds to the case of $\sigma_1\!=\!\sigma_4, \sigma_2\!=\!\sigma_3, m_1\!-\!\sigma_1=m_2\!-\!\sigma_2 (\equiv m_5)$ and $m_3 = m_4$.  In this case, the relevant matrix elements in eq.(\ref{Hexchf2}) combine to give
\begin{align}
&\sum_M
\langle m_5 \sigma_1|f_{m_3 \sigma_3}|(LS)JM \rangle
\langle (LS)JM|f_{m_3 \sigma_1}^\dagger |m_5 \sigma_3 \rangle
\nonumber \\
& = \frac{1}{2}
[1-(-1)^{3+3-L+\frac{1}{2}+\frac{1}{2}-S}]^2 (-1)^{\frac{1}{2}+\frac{1}{2}-S}
\nonumber \\
& \! \times
\sum_M \left[
\!\sum_{M_L,M_S}
\!\langle J M\!|\!L M_L,S M_S \rangle
\langle L M_L |3m_3,3m_5 \rangle
\!\langle S M_S|\frac{1}{2}\sigma_3, \frac{1}{2}\sigma_1 \rangle
\!\right]^2, \label{SpinInterchangeMatrix}
\end{align}
where we have used the following property of the 
Clebsch-Gordan coefficients: 
\begin{equation}
\langle j_2 m_2, j_1 m_1 | j_3 m_3 \rangle = (-1)^{j_1 + j_2 - j_3}
\langle j_1 m_1, j_2 m_2 | j_3 m_3 \rangle. \label{CGexchange}
\end{equation}
It is seen that $L+S$ must be even for eq.(\ref{SpinInterchangeMatrix}) to be nonzero. 
Since the Clebsch-Gordan coefficients are real, the right hand side of eq.(\ref{SpinInterchangeMatrix}) becomes positive 
for $S=1$, which is a property of $f_{\text{H}}^2$.
Considering that the energy denominator in eq.(\ref{Hexchf2}) is negative,
we derive a ferromagnetic on-site exchange, which 
is common with the double exchange mechanism.\cite{Zener}

In the case of the orbital exchange, we have a result similar to eq.(\ref{SpinInterchangeMatrix}) except that the sign factor is replaced by $(-1)^{3+3-L}$. 
With $S=1$, we obtain the negative sign factor, which brings about an antiferro-type interaction for the orbital exchange. 
The case for the simultaneous exchange of spin and orbit has the sign factor
$(-1)^{3+3-L+\frac{1}{2}+\frac{1}{2}-S}$ which is also negative for $f_{\text{H}}^2$.
Thus the corresponding interaction is again of antiferro-type.
The $4f^2$ multiplets have excited states with $S=0$.  With $S=0$ in the intermediate states we obtain different signs depending on types of the exchange.
Table~\ref{ConsequeceOfIntermediate} summarizes the signs for various intermediate states. 
We note that these consequences come from the antisymmetry of the 4$f^2$ wave function.
\begin{table}
\caption{Signs of exchange interactions between conduction and 4$f$ electrons with given intermediate states where
F and AF mean ferro- and antiferro-couplings, respectively.
The spin triplet is in the column $f^2(S=1)$, and the singlet in   
$f^2(S=0)$.  The multiplet $f_{\textrm H}^2$ is a special case of $f^2(S=1)$.
} \label{ConsequeceOfIntermediate}
    \begin{tabular}{cccc} \hline
    exchange & $f^0$ & $f^2(S=1)$ & $f^2(S=0)$ \\ \hline
    spin & AF & F & AF \\
    orbit & AF & AF & F \\
    spin and orbit & AF & AF & AF \\ \hline
    \end{tabular}
\end{table}

To show the symmetry in eqs.(\ref{Hexchf0}) and (\ref{Hexchf2}) more clearly,
we introduce the creation operator of the total angular momentum $j$ as
\begin{equation}
f_{jm}^\dagger = \sum_{m',\sigma}f_{m'\sigma}^\dagger \langle 3m', \frac{1}{2}\sigma | jm \rangle . \label{FermionOperatorProjection}
\end{equation}
Using the Wigner-Eckart theorem, the matrix element of $f_{jm}^\dagger$ is rewritten
by its reduced matrix element as
\begin{equation}
	\langle j_1 m_1 |f_{j m}^\dagger | 0\rangle =
	(-1)^{j_1- m_1}
	\begin{pmatrix}
		j_1\! & \!j\! & \!0 \\
		-m_1\! & \!m\! & \!0
	\end{pmatrix}
	\langle j_1 ||f_{j}^\dagger ||0 \rangle. \label{WignerEckart}
\end{equation}
Using this relation and introducing another angular momentum $P_1$ and the 6-$j$ symbol,\cite{Brink,Lindgren,Varshalovich} 
we write the part including the matrix elements of fermion operators as
\begin{align}
& \langle j_1 m_1 | f_{j_t m_t}^\dagger | 0\rangle
\langle 0| f_{j_u m_u} | j_1 m_2 \rangle \nonumber \\
&=- (-1)^{j_1 - m_1 + j_t - m_t}\sum_{P_1,Q_1}\Lambda(f^0;j_t,j_u,P_1) [P_1]
\begin{pmatrix}
		j_1\! & \!P_1\! & \!j_1 \\
		-m_1\! & \!Q_1\! & \!m_2
\end{pmatrix}
\begin{pmatrix}
		j_t\! & \!P_1\! & \!j_u \\
		-m_t\! & \!Q_1\! & \!m_u
\end{pmatrix},
\\
& \sum_{M}
\langle j_1 m_1|f_{j_u m_u}|(LS)JM \rangle
\langle (LS)JM|f_{j_t m_t}^\dagger |j_1 m_2 \rangle \nonumber \\
& = (-1)^{j_1 - m_1 + j_t - m_t}
\sum_{P_1,Q_1}\Lambda(f_{\textrm{H}}^2; j_t,j_u,P_1)[P_1]
\begin{pmatrix}
		j_1\! & \!P_1\! & \!j_1 \\
		-m_1\! & \!Q_1\! & \!m_2
\end{pmatrix}
\begin{pmatrix}
		j_t\! & \!P_1\! & \!j_u \\
		-m_t\! & \!Q_1\! & \!m_u
\end{pmatrix}, 
\end{align}
where $[j] \equiv 2j+1$ and 
\begin{align}
&\Lambda(f^0; j_t, j_u, P_1) \!=-
\langle j_1 ||f_{j_t}^\dagger ||0 \rangle
\langle j_1 ||f_{j_u}^\dagger ||0 \rangle^\ast
(-1)^{j_1 + 0 + j_u - P_1}
\begin{Bmatrix}
		j_1 & j_1 & P_1 \\
		j_t & j_u & 0
\end{Bmatrix}, \label{LambdaZero}
\\
&\Lambda(f_{\text{H}}^2; j_t, j_u, P_1) \!=
\langle (LS)J || f_{j_u}^\dagger ||j_1 \rangle^\ast
\langle (LS)J || f_{j_t}^\dagger ||j_1 \rangle
(-1)^{j_1 + j_t + J_2 + 2P_1}
\begin{Bmatrix}
		j_1 & j_1 & P_1 \\
		j_t & j_u & J
\end{Bmatrix}. \label{LambdaTwo}
\end{align}
The curly brackets in eqs.(\ref{LambdaZero}) and (\ref{LambdaTwo}) are the 6-$j$ symbols.
We note that $\Lambda(\xi;j,j',P)$ includes information of wave functions of the ground state and the excited state specified by $\xi$.
The transformation described above is best handled by a diagrammatic technique for addition of angular momenta.\cite{Brink,Lindgren}
Using the theory of angular momentum and the antisymmetry of wave functions,\cite{Lindgren}
we find that
\begin{align}
&\langle j'|| f_{j}^\dagger ||0 \rangle = -\delta_{j',j}[j']^{1/2},\\
&\langle (L S) J||f_{j}^\dagger||j' \rangle =(-1)^{j-j'+J}\sqrt{2}[L,S,j,j',J]^{1/2}
\begin{Bmatrix}
	3 & 3 & L \\
	1/2 & 1/2 & S \\
	j & j' & J
\end{Bmatrix}, \label{RMEf2}
\end{align}
where the curly bracket in eq.(\ref{RMEf2}) is the 9-$j$ symbol,\cite{Brink,Lindgren,Varshalovich}
and the shorthand notation
$[j_1,\cdots,j_n] = (2j_1+1)
\cdots (2j_n+1)$ is used.
In this way we obtain the following form:
\begin{align}
H_{\text{exch}}
& =\frac{4\pi}{N}|V_{k_F}|^2
\sum_{i,\mib{k},\mib{k}'}
\sum_{j_t,j_u}^{5/2,7/2}
\sum_{P_1 = 0}^{2j_1}
\sum_{\xi_i}^{f^0,f_{\text{H}}^2}
[P_1]
\frac{\Lambda(\xi_i; j_t,j_u,P_1)}{E_{\text{exc}}(\xi_i)}
\nonumber \\
& \times
\sum_{\sigma, \sigma'}
\sum_{m_t,m_u}
Y_{3m_t - \sigma}^\ast (\Omega_{\mib{k}})
Y_{3m_u - \sigma'} (\Omega_{\mib{k}'})e^{i(\mib{k} - \mib{k}') \cdot \mib{R}_i}
\nonumber \\
& \times
(-1)^{j_1 - m_1+j_t - m_t}
\langle j_t m_t | 3m_t - \sigma, \frac{1}{2} \sigma \rangle
\langle 3m_u - \sigma', \frac{1}{2}\sigma' | j_u m_u \rangle
\nonumber \\
& \times
\sum_{m_1,m_2}
\sum_{Q_1= -P_1}^{P_1}
\begin{pmatrix}
		j_1 & P_1 & j_1 \\
		-m_1 & Q_1 & m_2
\end{pmatrix}
\begin{pmatrix}
		j_t & P_1 & j_u \\
		-m_t & Q_1 & m_u
\end{pmatrix}
\nonumber \\
& \times
|i; j_1 m_1 \rangle \langle i; j_1 m_2 |
c_{\mib{k}' \sigma'}^\dagger c_{\mib{k} \sigma}, \label{Hexchfinal}
\end{align}
where $\xi_i$ is the index of intermediate states at $\mib{R}_i$, which is either $f^0$ or $f_{\text{H}}^2$.
The excitation energy
$E_{\text{exc}}$ is defined as follows:
\begin{align}
&E_{\text{exc}}(f^0) = E_1 - E_0 - \epsilon_F, \\
&E_{\text{exc}}(f_{\text{H}}^2) = E_1 - E_2 + \epsilon_F.
\end{align}

In place of the set of $(2j_1 + 1)^2$ operators $|j_1 m_1 \rangle \langle j_1 m_2 |$ with $ m_1, m_2 = -j_1,\ldots, j_1 $, we can use more conveniently the set of irreducible tensor operators $T_Q^P$, with $ Q = -P,\ldots,P $, and $ P = 0, \ldots ,2j_1$.\cite{Judd,Teitelbaum}
The irreducible tensor operator is related to
$ |i;j_1 m_1 \rangle \langle i; j_1 m_2 |$ by,\cite{Teitelbaum}
\begin{align}
|i;j_1 m_1 \rangle \langle i; j_1 m_2| &= (-1)^{j_1-m_1}\sum_{P = 0}^{2j_1}\sum_{Q = -P}^P
\frac{[P]}{\langle j_1|| \mib{T}^P || j_1 \rangle}
\begin{pmatrix}
		\!j_1\! & \!P\! & \!j_1\! \\
		\!-m_1\! & \!Q\! & \!m_2\!
\end{pmatrix}
T_Q^P(i), \label{DefITO2}
\end{align}
where 
$\langle j_1 ||T^P||j_1 \rangle$ ($P = 1, \ldots, 2j_1$) is a reduced matrix element.

By inserting eq.(\ref{DefITO2}) into eq.(\ref{Hexchfinal}),
we obtain another expression for the on-site exchange Hamiltonian 
\begin{align}
H_{\text{exch}}
& =\frac{4\pi}{N}|V_{k_F}|^2
\sum_{i,\mib{k},\mib{k}'}
\sum_{j_t,j_u}^{5/2,7/2}
\sum_{P_1 = 0}^{2j_1}
\sum_{\xi_i}^{f^0,f_{\text{H}}^2}
[P_1]
\frac{\Lambda(\xi_i; j_t,j_u,P_1)}{E_{\text{exc}}(\xi_i)\langle j_1 || T^{P_1} || j_1 \rangle}
\nonumber \\
& \times
\sum_{\sigma, \sigma'}
\sum_{m_t,m_u}
Y_{3m_t - \sigma}^\ast (\Omega_{\mib{k}})
Y_{3m_u - \sigma'} (\Omega_{\mib{k}'})e^{i(\mib{k} - \mib{k}') \cdot \mib{R}_i}
\nonumber \\
& \times
\langle j_t m_t | 3m_t - \sigma, \frac{1}{2} \sigma \rangle
\langle 3m_u - \sigma', \frac{1}{2}\sigma' | j_u m_u \rangle
\nonumber \\
& \times
\sum_{Q_1= -P_1}^{P_1}
(-1)^{j_t - m_t}
\begin{pmatrix}
		j_t & P_1 & j_u \\
		-m_t & Q_1 & m_u
\end{pmatrix}
T_{Q_1}^{P_1}(i)
c_{\mib{k}' \sigma'}^\dagger c_{\mib{k} \sigma}, \label{HexchITO}
\end{align}
where the orthogonality relation of 3-$j$ symbols is used.\cite{Lindgren}

\subsection{Effective intersite interaction}

In order to derive the intersite interaction we work with  the waver-number representation of the on-site interaction: 
\begin{align}
H_{\text{exch}} =
&
\sum_{i,\xi_i,P_i, Q_i}
\sum_{\mib{k}, \mib{k}', \sigma, \sigma'}
W_{\mib{k}' \sigma' \mib{k} \sigma}^{P_i Q_i}(\xi_i)
e^{i(\mib{k}-\mib{k}') \cdot \mib{R}_i} T_{Q_i}^{P_i}(i)
c_{\mib{k}'\sigma'}^\dagger c_{\mib{k}\sigma}, \label{Hkf}
\end{align}
where 
the matrix element
$W_{\mib{k}' \sigma' \mib{k} \sigma}^{P_i Q_i}(\xi_i)$
is determined by comparison with eq.(\ref{HexchITO}).
Then the effective intersite interaction between two Ce ions, which are located at $\mib{R}_1$ and $\mib{R}_2$, is given by the second-order perturbation theory with respect to $H_{\text{exch}}$ as
\begin{align}
H_{12} =& \frac{1}{2}\sum_{\mib{k}, \mib{k}', \sigma, \sigma'}
\frac{f(\epsilon_k)-f(\epsilon_{k'})}{\epsilon_k - \epsilon_k'}
e^{i(\mib{k}-\mib{k}') \cdot (\mib{R}_1 - \mib{R}_2)}
\sum_{P_1, P_2, Q_1, Q_2}
T_{Q_1}^{P_1}(1) \otimes T_{Q_2}^{P_2}(2) 
\nonumber \\
& \times
\sum_{\xi_1,\xi_2}
W_{\mib{k}' \sigma' \mib{k} \sigma}^{P_1 Q_1}(\xi_1)
W_{\mib{k} \sigma \mib{k}' \sigma'}^{P_2 Q_2}(\xi_2)
. \label{PAIRHAMILTONIAN}
\end{align}
We replace the sum over $\mib{k}$ by integrals,
\begin{equation}
\sum_{\mib{k}} \rightarrow N \left(\frac{a}{2\pi}\right)^3 \int_0^\infty \, k^2 dk \, \int \, d\Omega_{\mib{k}}.
\end{equation}
The angular integral is of the form
\begin{equation}
\int d\Omega_{\mib{k}}
Y_{3m_f}^\ast (\Omega_{\mib{k}})Y_{3m_i}(\Omega_{\mib{k}})
\exp(\pm i\mib{k}\cdot \mib{R}),
\end{equation}
for which we use the partial-wave expansion of a plane wave:
\begin{equation}
\exp(\pm i\mib{k}\cdot\mib{r})=4\pi \sum_{l}(\pm i)^l j_l(kr)
\sum_{m}Y_{lm}^\ast(\Omega_{\mib{k}})Y_{lm}(\hat{r}), \label{PWE}
\end{equation}
where $\hat{r}\equiv\mib r/r$ denotes the unit vector in the direction of $\mib r$.
Furthermore  the decomposition of the product of spherical harmonics with the same argument:\cite{Lindgren}
\begin{align}
Y_{l_1 m_1}(\Omega)Y_{l_2 m_2}(\Omega)
&= \frac{1}{\sqrt{4\pi}}\sum_{L = 0}^\infty
(-1)^L [L, l_1, l_2]^{\frac{1}{2}}
\begin{pmatrix}
		\!L\! & \!l_1\! & \!l_2\! \\
		\!0\! & \!0\! & \!0\!
\end{pmatrix}
\nonumber \\
& \times
\sum_{M = -L}^L
(-1)^{L-M}
\begin{pmatrix}
	\!L\! & \!l_1\! & \!l_2\! \\
	\!-M\! & \!m_1\! & \!m_2\!
\end{pmatrix}
Y_{LM}(\Omega).
\end{align}
Then we obtain
\begin{align}
& \int \, d\Omega_{\mib{k}}
Y_{3m_f}^\ast(\Omega_{\mib{k}})Y_{3m_i}(\Omega_{\mib{k}})
\exp(\pm i \mib{k} \cdot \mib{R}) \nonumber \\
& =  (-1)^{m_i}\sqrt{4\pi}[3]\sum_{l = 0}^\infty (\pm i)^l j_l(kR)
[l]^{\frac{1}{2}}
\begin{pmatrix}
	3\! & \!l\! & \!3\! \\
	0\! & \!0\! & \!0\!
\end{pmatrix}
\sum_{m = -l}^l
\begin{pmatrix}
	\!3\! & \!\!l\!\! & \!3\! \\
	\!m_f\! & \!\!m\!\! & \!-m_i\!
\end{pmatrix}
Y_{lm} (\hat{R}). \label{dOmegaK}
\end{align}
In eq.(\ref{dOmegaK}) the first 3-$j$ symbol vanishes for odd $l$,\cite{Lindgren} hence 
we have to consider even $l$ only.

In the next stage we take summation over intermediate angular momenta, and simplify the results by using symmetry properties. 
For example, the summation over $\sigma,\sigma'$ in eq.(\ref{HexchITO}) gives rise to a 6-$j$ symbol:\cite{Brink,Lindgren}
\begin{align}
&\sum_{\sigma,\sigma'}
\langle j_t m_t | 3m_t-\sigma, \frac{1}{2} \sigma \rangle
\langle 3m_s - \sigma, \frac{1}{2} \sigma | j_s m_s \rangle
(-1)^{3+(m_s-\sigma)}
\begin{pmatrix}
		\!3\! & \!l_1\! & \!3\! \\
		\!{m_t\!-\!\sigma}\! & \!m\! & \!{-(m_s\!-\!\sigma)}\!
\end{pmatrix}
\nonumber \\
& =(-1)^{-j_s + 3 + \frac{1}{2}}
[j_s,j_t]^{1/2}
\left\{
	\begin{array}{ccc}
		\!3\! & \!3\! & \!l_1\! \\
		\!j_t\! & \!j_s\! & \!\frac{1}{2}\!
	\end{array}
\right\}
(-1)^{j_s - m_s}
\left(\!
	\begin{array}{ccc}
		\!j_t\! & \!l_1\! & \!j_s\! \\
		\!m_t\! & \!m\! & \!-m_s\!
	\end{array}
\!\right), \label{JMJM}
\end{align}
and the summation over internal azimuthal quantum numbers yields
the 9-$j$ symbol:\cite{Brink,Lindgren} 
\begin{align}
&\sum_{m_s, m_t, m_u, m_v, m, m'}
(-1)^{({j_t - m_t})+({j_s - m_s})+({j_u - m_u})+({j_v - m_v})+({L - M})}
\begin{pmatrix}
		\!j_t\! & \!P_1\! & \!j_u\! \\
		\!-m_t\! & \!Q_1\! & m_u\!
\end{pmatrix}
\nonumber \\
& \times
\begin{pmatrix}
		\!j_t\! & \!l_1\! & \!\!j_s\! \\
		\!m_t\! & \!m\! & \!\!-m_s\!
\end{pmatrix}
\begin{pmatrix}
		\!j_v\! & \!l_2\! & \!\!j_u\! \\
		\!m_v\! & \!m'\! & \!\!-m_u\!
\end{pmatrix}
\begin{pmatrix}
		\!j_v\! & \!P_2\! & \!j_s\! \\
		\!-m_v\! & \!Q_2\! & \!m_s\!
\end{pmatrix}
\begin{pmatrix}
		\!L\! & \!l_1\! & \!l_2\! \\
		\!-M\! & \!m\! & \!m'\!
\end{pmatrix}
\nonumber \\
& =(-1)^{P_2 + L +l_2 + j_t - j_v}
\begin{Bmatrix}
		\!P_1\! & \!j_u\! & \!j_t\! \\
		\!P_2\! & \!j_v\! & \!j_s\! \\
		\!L\! & \!l_2\! & \!l_1\!
\end{Bmatrix}
\begin{pmatrix}
		\!P_1\! & \!P_2\! & \!L\! \\
		\!Q_1\! & \!Q_2\! & \!M\!
\end{pmatrix}.
\end{align}
We then take the
integral over radial momenta $k, k'$.   This is given by
\begin{align}
Q_{l_1 l_2}(k_F R)
= (k_F R)^{-4} \!\int_0^{k_F R} x^2 dx \int_0^\infty 
\frac{{x'}^2 dx'}{x^2 - {x'}^2}
\left\{
	j_{l_1}(x)j_{l_2}(x') + j_{l_1}(x')j_{l_2}(x)
\right\},
\label{radialQ}
\end{align}
where $j_l(x)$ is the spherical Bessel function of order $l$ and the energy dispersion for the free electron $\epsilon_k = \hbar^2 k^2/2m$ is inserted.
Note that $Q_{l_1l_2}(k_FR)$ is invariant under interchange of $l_1$ and $l_2$ 
by its definition.
Appendix gives the analytic results for $Q_{l_1 l_2}(k_F R)$ for all relevant values of $l_1 l_2$. 
Putting all together we finally obtain the 
following form:
\begin{align}
&H_{12} =
\frac{|V_{k_F}|^4}{4\pi^{\frac{7}{2}}E_a}
\sum_{\xi_1,\xi_2}
\sum_{l_1, l_2, L = 0}^\infty
\sum_{P_1, P_2 = 0}^{2j_1}
\sum_{j_s, j_t, j_u, j_v}^{\frac{5}{2}, \frac{7}{2}} \nonumber \\
& \times
\frac{
\Lambda(\xi_1; j_t,j_u,P_1)\Lambda(\xi_2; j_v, j_s, P_2 )
}{E_{\text{exc}}(\xi_1)E_{\text{exc}}(\xi_2)}
(k_F a)^4Q_{l_1 l_2}(k_F R_{12})
\nonumber \\
& \times
(-1)^{(l_1+l_2)/2-j_s +j_t -j_u - j_v - P_1 + 1}
[3]^2 [l_1, l_2, P_1, P_2][j_s, j_t, j_u, j_v, L]^{\frac{1}{2}}
\nonumber \\
& \times
\begin{pmatrix}
		\!3\! & \!l_1\! & \!3\! \\
		\!0\! & \!0\! & \!0\!
\end{pmatrix}
\begin{pmatrix}
		\!3\! & \!l_2\! & \!3\! \\
		\!0 & \!0\! & \!0\!
\end{pmatrix}
\begin{pmatrix}
		\!L\! & \!l_1\! & \!l_2\! \\
		\!0\! & \!0\! & \!0\!
\end{pmatrix}
\begin{Bmatrix}
		\!3\! & \!3\! & \!l_1\! \\
		\!j_t\! & \!j_s\! & \frac{1}{2}
\end{Bmatrix}
\begin{Bmatrix}
		\!3\! & \!3\! & \!l_2\! \\
		\!j_v\! & \!j_u\! & \!\frac{1}{2}\!
\end{Bmatrix}
\begin{Bmatrix}
		\!P_1\! & \!j_u\! & \!j_t\! \\
		\!P_2\! & \!j_v\! & \!j_s\! \\
		\!L\! & \!l_2\! & \!l_1\!
\end{Bmatrix}
\nonumber \\
& \times
\frac{
\{\mib{T}^{P_1}(1)\mib{T}^{P_2}(2)\mib{Y}_L(\hat{R}_{12})\}_0^0
}{\langle j_1 || \mib{T}^{P_1} || j_1 \rangle \langle j_1 || \mib{T}^{P_2} || j_1 \rangle}, \label{H12ITO}
\end{align}
where $\mib{R}_{12} = \mib{R}_1-\mib{R}_2$.
In eq.(\ref{H12ITO}) we have used the fact that $L, l_1, l_2$ are even integers, and 
introduced $E_a=\hbar^2/2ma^2$.
In eq.(\ref{H12ITO}), $\{ \cdots \}_0^0$ denotes a tensor component of rank zero, i.e., a scalar,\cite{Teitelbaum,Lindgren}  It can also be written as
\begin{align}
&\{\mib{T}^{P_1}(1)\mib{T}^{P_2}(2)\mib{Y}_L(\hat{R}_{12})\}_0^0\nonumber \\
&= (-1)^{P_1+P_2+L}
\sum_{Q_1,Q_2,M}
\begin{pmatrix}
	\!P_1\! & \!P_2\! & \!L\! \\
	\!Q_1\! & \!Q_2\! & \!M\!
\end{pmatrix}
T_{Q_1}^{P_1}(1)T_{Q_2}^{P_2}(2)Y_{LM}(\hat{R}_{12}).
\label{tensorproduct}
\end{align}

Among various components of the Hamiltonian,
those terms with $L=0$ represents 
the isotropic interactions.
Other terms with $L\neq 0$ represents anisotropic interactions
which depend on the relative angular position of the two ions.
For example, the pseudo-dipole interaction
corresponds to the case
$P_1= P_2=1$ and $L=2$.\cite{Kasuya}
Let us choose $\mib{R}_{12}$ along the quantization axis $z$. 
Then we have $\hat{R}_{12}=\hat{z}$, and  
$Y_{LM}(\hat{R}_{12})$ vanishes except for $M=0$.
Then the dipole interaction has the following form
\begin{equation}
K(\mib R_{12})[3j_z (1)j_z (2) - \mib{j}(1)\cdot \mib{j}(2)].
\label{dipolar}
\end{equation}
With $K(\mib R_{12})<0$,  this interaction favors parallel moments along the $z$ axis.
For moments directed perpendicular to the $z$ axis, the first term of eq.(\ref{dipolar}) vanishes and the second term favors antiparallel moments.
This preference is just as in the case of real dipoles.
On the other hand, with $K(\mib R_{12})>0$, moments along $\hat{z}$
 tends to be antiparallel, and those perpendicular to $\hat{z}$ tends to be parallel.
We refer to the case of $K(\mib R_{12})>0$ as the ``anti-dipole" interaction.

\subsection{Projection to crystal-field ground states}

So far we have not included any element of the crystal structure.   
In CeB$_6$, the cubic symmetry around each trivalent Ce ion splits the sixfold degenerate ground state with $j_1 = 5/2$ into a quartet called $\Gamma_8$, and a doublet called $\Gamma_7$.
The crystalline electric field (CEF) splitting between $\Gamma_8$ and $\Gamma_7$
is about 540 K.\cite{Zirngiebl} 
Hence at temperatures ($< 10$K) of our interest we can safely neglect the population in the $\Gamma_7$ excited states.
In considering the intersite interactions, we therefore pick out such component of $4f$ wave functions that belong to the $\Gamma_8$ states.  
In other words we perform 
projection of $|j_1 m \rangle$ to $\Gamma_8$ wave functions $|\nu \sigma \rangle$ ($\nu = +,-$ and $\sigma = \uparrow, \downarrow$) by
\begin{equation}
|j_1 m \rangle 
\rightarrow 
\sum_{\nu, \sigma} | \nu \sigma\rangle \langle \nu \sigma| m \rangle . \label{projection2CEF}
\end{equation}
where $\nu$ specifies the orbital index and $\sigma$ the Kramers index, i.e., one of  the time-reversal partners.
We have omitted writing $j_1=5/2$ in the kets $|j_1 m \rangle$ of the right-hand side. 
Explicitly the 
$\Gamma_8$ quartet has the following wave functions:
\begin{align}
| + \uparrow \rangle &= \sqrt{\frac{5}{6}} \left| +\frac{5}{2} \right\rangle
+ \sqrt{\frac{1}{6}}\left| -\frac{3}{2} \right\rangle , \\
| + \downarrow \rangle &= \sqrt{\frac{5}{6}} \left| -\frac{5}{2} \right\rangle
+ \sqrt{\frac{1}{6}}\left| +\frac{3}{2} \right\rangle ,\\
| - \uparrow \rangle &= \left| +\frac{1}{2} \right\rangle , \\
| - \downarrow \rangle &= \left| -\frac{1}{2} \right\rangle ,
\end{align}
where each coefficient in front of  $|m\rangle$ corresponds to $\langle m| \nu \sigma \rangle (=\langle \nu \sigma| m \rangle)$.

The transitions within the $\Gamma_8$ have $15 (=4\times 4-1)$ components, and
the operators describing these transitions correspond to 15 multipole operators.
Following the literature,\cite{Khomskii,Ohkawa} we introduce two pseudo-spins $\mib\sigma$ and $\mib\tau$ in terms of   
the vector $\mib{\rho}_{\tau, \tau'}$ of the Pauli matrices:
\begin{align}
\mib{\tau} &= \sum_{\sigma,\tau,\tau'}
| \tau \sigma \rangle \mib{\rho}_{\tau, \tau'} \langle \tau' \sigma | ,\\
\mib{\sigma} &= \sum_{\sigma, \sigma', \tau}
| \tau \sigma \rangle \mib{\rho}_{\sigma, \sigma'} \langle \tau \sigma' | ,
\end{align}
Using these pseudo-spins, we can express a physical operator $X^{\gamma}$ adapted to the point group symmetry,\cite{Shiina,Kusunose}
Here $\gamma$ specifies a component in the irreducible representation $\Gamma$ with possible multiplicity. 
\begin{table}
\caption{The list of multipole operators in the $\Gamma_8$ subspace.} \label{MOG8}
		\begin{tabular}{
		ccc}	\hline
		$\Gamma$ \qquad & symmetry \qquad & $X^{\gamma}$ \\ \hline
		2$u$ 	& $\sqrt{15}xyz$ &			$\tau^y$ \\
		3$g$ 	& $(3z^2-r^2)/2$ &			$\tau^z$ \\
		\, 		& $\sqrt{3}(x^2-y^2)/2$ &	$\tau^x$ \\
		4$u$1	& $x$	& 					$\sigma^x$ \\
		\,		& $y$	& 					$\sigma^y$ \\
		\,		& $z$	& 					$\sigma^z$ \\
		4$u$2	& $x(5x^2-3r^2)/2$	& 		$\eta^x$ \\
		\,		& $y(5y^2-3r^2)/2$	& 		$\eta^y$ \\
		\,		& $z(5z^2-3r^2)/2$	& 		$\eta^z$ \\
		5$u$	& $\sqrt{15}x(y^2-z^2)/2$	& $\zeta^x$ \\
		\,		& $\sqrt{15}y(z^2-x^2)/2$	& $\zeta^y$ \\
		\,		& $\sqrt{15}z(x^2-y^2)/2$	& $\zeta^z$ \\
		5$g$	& $\sqrt{3}yz$				& $\mu^x$ \\
		\,		& $\sqrt{3}zx$				& $\mu^y$ \\
		\,		& $\sqrt{3}xy$				& $\mu^z$ \\ \hline
		\end{tabular}
\end{table}
For completeness the multipole operators are summarized in Table~\ref{MOG8}, where 
$\mib{\mu},\mib{\eta}, \mib{\zeta}$ are defined by
\begin{align}
&\mib{\mu} =(\mu^x, \mu^y, \mu^z) = (\tau^y \sigma^x, \tau^y \sigma^y, \tau^y \sigma^z) , \\
&\mib{\eta} =(\eta^x, \eta^y, \eta^z) 
= \frac{1}{2}(-\tau^z \sigma^x \!+\! \sqrt{3}\tau^x \sigma^x,
							-\tau^z \sigma^y \!-\! \sqrt{3}\tau^x \sigma^y,
							2\tau^z \sigma^z), \\
&\mib{\zeta} =(\zeta^x, \zeta^y, \zeta^z) 
= \frac{1}{2}(-\sqrt{3}\tau^z \sigma^x \!-\! \tau^x \sigma^x,
								\sqrt{3}\tau^z \sigma^y \!-\! \tau^x \sigma^y,
								2\tau^x \sigma^z). 
\end{align}
The subscript $u$ represents the odd property under the time reversal, and $g$ the even one.
The operators belonging to $\Gamma_{3g}$ and $\Gamma_{5g}$ are classified
into quadrupole operators.
Under the cubic symmetry the dipole operator $\mib{J}$ are mixed with octupole operators.
Hence 
$\mib{\sigma}$ and $\mib{\eta}$ have both dipole and octupole characters.
The remaining representations $\Gamma_{2u}$ and $\Gamma_{5u}$ describe pure octupole operators.
In Table~\ref{TandX}, the relations between irreducible tensor operators $T_Q^P$ and physical operators $X^{\gamma}$ are summarized.\cite{TandXComment}
\begin{table}
\caption{Projection from $T_Q^P$ into $X^{\gamma}$.} \label{TandX}
	\begin{tabular}{ccc} \hline
	$P$ & $Q$ & \qquad $T_Q^P/\langle j_1||T^P||j_1 \rangle$ \\ \hline
	1 & $\pm1$ &
		$\frac{1}{6\sqrt{105}}\left[
		4(\mp \eta^x - i \eta^y)
		+7(\mp \sigma^x - i\sigma^y)
		\right]$ \\
	1 & $0$ &
		$\frac{1}{6}\sqrt{\frac{2}{105}}
		\left(4\eta^z + 7\sigma^z \right)$ \\ \hline
	2 & $\pm 2$ &
		$\frac{1}{2\sqrt{210}}
		\left( \pm i\mu^z+4\tau^x \right)$ \\
	2 & $\pm 1$ &
		$\frac{1}{2\sqrt{210}}
		\left( -i\mu^x \mp \mu^y \right)$ \\
	2 & $0$ &
		$\frac{2}{\sqrt{105}}\tau^z$ \\ \hline
	3 & $\pm 3$ &
		$\frac{1}{36\sqrt{7}}
		\left[
		3\sqrt{3}(\pm \zeta^x+i\zeta^y)
		+7(\mp \eta^x + i\eta^y)
		+(\mp \sigma^x + i\sigma^y)
		\right]$ \\
	3 & $\pm 2$ &
		$\frac{1}{6\sqrt{14}}
		\left(
		2\zeta^z \pm 3i\tau^y
		\right)$ \\
	3 & $\pm 1$ &
		$\frac{1}{36\sqrt{35}}
		\left[
		15(\pm \zeta^x - i\zeta^y)
		+7\sqrt{3}(\pm \eta^x + i\eta^y)
		+\sqrt{3}(\pm \sigma^x + i\sigma^y)
		\right]$ \\
	3 & $0$ &
		$\frac{1}{9\sqrt{35}}
		\left(
		7\eta^z+\sigma^z
		\right)$ \\ \hline
	4 & $\pm 4$ &
		$\frac{1}{36}\sqrt{\frac{5}{2}}(1+\tau^z)$ \\
	4 & $\pm 3$ &
		$\frac{1}{12}\sqrt{\frac{5}{3}}(-i\mu^x \pm \mu^y)$ \\
	4 & $\pm 2$ &
		$\frac{1}{6}\sqrt{\frac{5}{42}}(\pm 2i \mu^z+\tau^x)$ \\
	4 & $\pm 1$ &
		$\frac{1}{12}\sqrt{\frac{5}{21}}(i \mu^x \pm \mu^y)$ \\
	4 & $0$ &
		$\frac{1}{36\sqrt{7}}(7-5\tau^z)$ \\ \hline
	5 & $\pm 5$ &
		$\frac{5}{48\sqrt{11}}
		\left[ \sqrt{3}(\pm \zeta^x - i\zeta^y)
		+ (\pm \eta^x +i \eta^y) + 2(\mp \sigma^x -i\sigma^y)\right]$ \\
	5 & $\pm 4$ &
		$\frac{1}{12}{\sqrt{\frac{5}{22}}}\left( \eta^z + \sigma^z \right)$ \\
	5 & $\pm 3$ &
		$\frac{1}{144} {\sqrt{\frac{5}{11}}}
		\left[ 3\sqrt{3}(\pm \zeta^x + i\zeta^y)
		+13(\mp \eta^x + i\eta^y)
		+ 2(\pm \sigma^x - i\sigma^y) \right]$ \\
	5 & $\pm 2$ &
		$\frac{1}{6}{\sqrt{\frac{5}{22}}}\zeta^z$ \\
	5 & $\pm 1$ &
		$\frac{1}{72}{\sqrt{\frac{5}{154}}}
	    \left[21(\mp \zeta^x + i \zeta^y)
	    +\sqrt{3}(\pm \eta^x + i \eta^y)
	    +14\sqrt{3}(\mp \sigma^x - i \sigma^y)\right]$ \\
	5 & $0$ &
		$\frac{5}{36\sqrt{77}}\left(-5\eta ^z + 7\sigma ^z \right)$ \\ \hline
	\end{tabular}
\end{table}
As seen in the last section, the effective Hamiltonian (\ref{H12ITO}) involves the product of operators with different ranks for each site.
Hence after projection of the Hamiltonian into the $\Gamma_8$ states, there should be  couplings between different irreducible representations.
Symmetry analysis of such couplings has been performed recently by Sakai et al. in analogy to the Slater-Koster scheme for the energy band theory.\cite{SakaiN}
In our case, the general discussion involves very many terms and is not illuminating.
Therefore we focus in this paper only on the couplings within each irreducible representation $\Gamma$.

In analogy to the dipole interaction given by eq.(\ref{dipolar}), we refer to the interactions of the form:
\begin{equation}
H_{\rm pseudo} = K(\mib R_{12})  
\{3
[\mib v(\mib R_1)\cdot\hat{R}_{12}]
[\mib v(\mib R_2)\cdot\hat{R}_{12}]
- \mib v(\mib R_1)\cdot \mib v(\mib R_2)\},
\label{pseudo-dipole}
\end{equation}
as the pseudo-dipole interaction in the $\Gamma_8$ subspace. 
Here $\mib v$ is either of $\mib\sigma, \mib\eta, \mib\zeta$ in Table~\ref{MOG8}.
The case $K(\mib R_{12}) >0$ is called the ``anti-pseudo-dipole" interaction.

\section{Explicit Results for Multipolar Interactions}

In this section we derive strength and spatial dependence 
of multipolar interaction for a pair of Ce ions 
separated by $\mib{R}$.
We assume that the Fermi surface consists of a sphere with radius $k_F$.
Assuming the same number of electrons as that of the lattice sites with the spacing $a$,
we obtain $k_F a= (3\pi^2)^{1/3} \sim 3.09$.
Let $R_{nn}$ represent the distance between the nearest-neighbor pair,
and $R_{nnn}$ between the next-nearest-neighbor pair.
Then we have $k_F R_{nn}\sim 3.09$ and $k_F R_{nnn}\sim 4.37$ for the simple cubic lattice. 
We note that $a =4.14\AA$ in CeB$_6$.
For evaluation of quadrupolar couplings, we take $\mib{R}$ along the (001) direction because couplings between nearest neighbors are important for AFQ ordering in CeB$_6$.
When $\mib{R}$ is parallel to 
the $z$-axis, $Y_{LM}(\hat{R})$ vanishes unless $M=0$ in eq.(\ref{tensorproduct}).
This gives a high symmetry among components of multipole interactions.
On the other hand, $Y_{LM}(\hat{R})$ with $M \neq 0$ does not vanish for the directions of next-nearest neighbors.  This gives a chance of larger anisotropy to the next-nearest-neighbor interaction.
For the rest of this section, we define three cases depending on intermediate states on two sites:
(a) 4$f^0$ intermediate states on both sites; (b) 4$f_{\textrm{H}}^2$ intermediate states on one site and 4$f^0$ on the other site; (c) 4$f_{\text{H}}^2$ intermediate states on both sites.

We comment on the relationship between the present model and the actual electronic structure of CeB$_6$.
The real Fermi surface of CeB$_6$ consists of three equivalent sheets, which are nearly spherical. It has been shown that not only the polarization within each pocket centered at $X$ points in the Brillouin zone, 
but also the interpocket polarization plays an important role in determining the Ruderman-Kittel-Kasuya-Yosida (RKKY) interaction.\cite{kuramoto-kubo}
Thus, taking a single Fermi surface our model incorporates 
both intra- and inter-pocket polarizations in a very rough manner.
Thus in a sense a model with the single Fermi surface is better than considering only the intrapocket polarization of three Fermi surfaces. 
In relation to the  magnetic structures observed in phases II and III, 
we are particularly interested in the coupling constant for nearest and next-nearest neighbors.

\subsection{Quadrupole-quadrupole interaction}
We begin with 
the quadrupolar interactions, paying attention to the 
strength and sign at the distance between the nearest-neighbor pair.
Namely, we are interested in whether the simple hybridization model can explain the antiferro-quadrupole order of the $\Gamma_5$ type.
We define the quadrupole-quadrupole interaction Hamiltonian as follows:
\begin{align}
H_Q &= \frac{|V_{k_F}|^4}{E_a} \sum_{\xi_1,\xi_2}
\frac{1}{E_{\textrm{exc}}(\xi_1)E_{\textrm{exc}}(\xi_2)} \nonumber \\
& \times[
\sum_{\alpha,\beta}^{z,x}D_{3g}^{\alpha \beta}(\xi_1, \xi_2; \mib  R) \tau^\alpha (1)\tau^\beta (2)
+\sum_{\alpha,\beta}^{x,y,z}D_{5g}^{\alpha \beta} (\xi_1, \xi_2; \mib  R) \mu^\alpha (1)\mu^\beta (2)
],
\end{align}
where $\tau^\alpha$ and $\mu^\alpha$ have been defined in Table~\ref{MOG8}, and
$\xi_1$ and $\xi_2$ specify 
either $4f^0$ or $4f_{\text H}^2$ as intermediate states 
at $\mib{R}_1$ and $\mib{R}_2$, respectively.
The dimensionless quantities $D_{\Gamma}^{\alpha \beta}(\xi_1, \xi_2; \mib  R)$ with $\Gamma$ being $3g$ or $5g$ are determined by eq.(\ref{H12ITO}) together with the projection to $\Gamma_8$.
We use the matrix notation $\mib{D}_{\Gamma}$ to represent the set of components.
The form of the intersite interaction depends on the excitation energies $E_{\textrm{exc}}(\xi)$, since the $\mib R$-dependence of $\mib{D}_{\Gamma}(\xi_1,\xi_2;\mib  R)$ varies depending on $\xi_1$ and $\xi_2$.
Namely, if the Ce ion tends to become tetravalent, we obtain $|E_{\textrm{exc}}(f^0)| < |E_{\textrm{exc}}(f^2)|$ and the 
$f^0$ intermediate state dominates.
On the other hand, in the case of comparable $E_{\textrm{exc}}(f^0)$ and 
$ E_{\textrm{exc}}(f^2)$, all 
$\mib{D}_{\Gamma}(\xi_1,\xi_2;\mib  R)$ also have comparable multiplying factor. The situation in CeB$_6$ seems to correspond to the latter. 

\begin{figure}[htbp]
\begin{center}
	\includegraphics[width=8cm, clip]{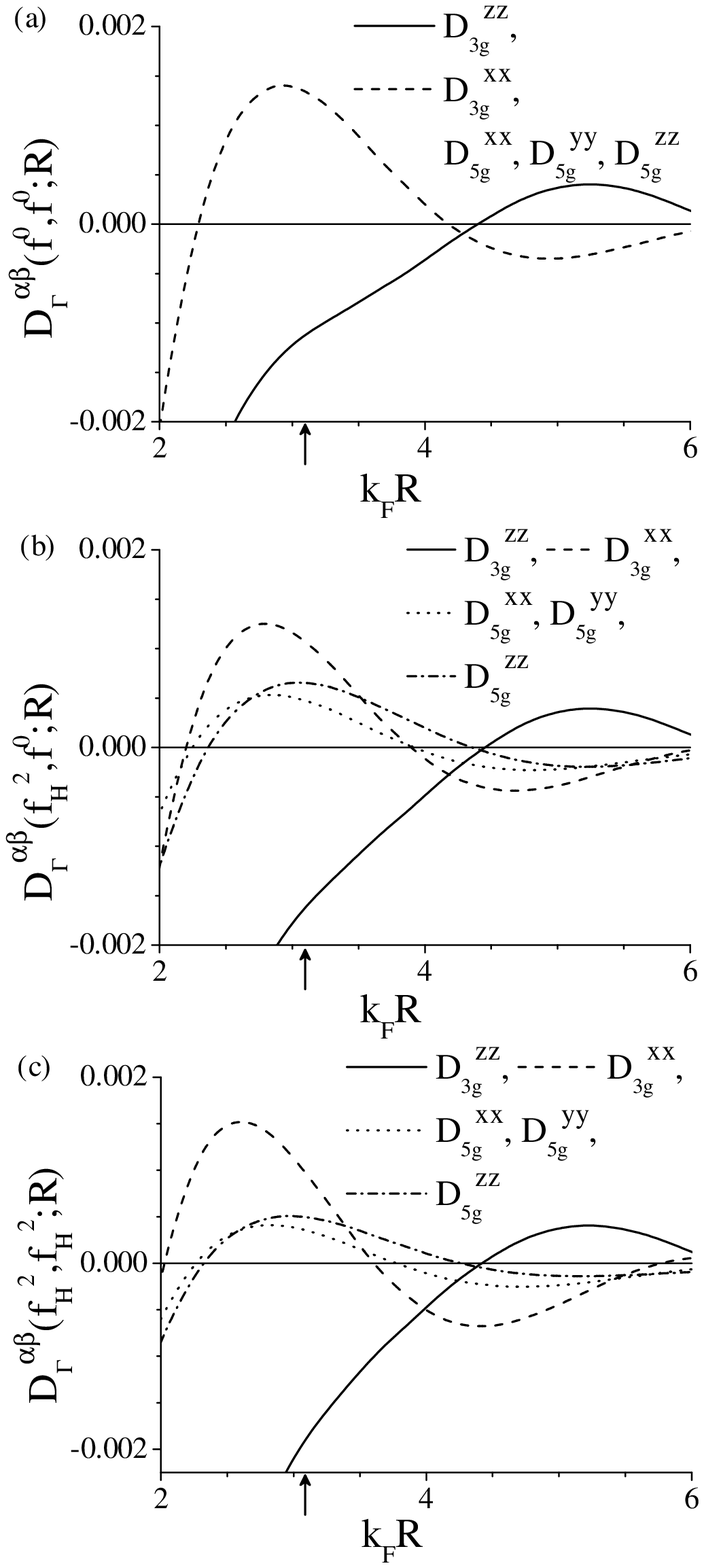}
\end{center}
\caption{
Spatial dependences of quadrupole-quadrupole interactions $\mib{D}_{3g}$ and $\mib{D}_{5g}$ with 
$\mib{R}=(0,0,R)$.
Results with different intermediate states are shown as follows: 
(a) $f^0$ intermediate states on both sites;
(b) $f_{\textrm{H}}^2$ intermediate states on one site and $f^0$ on the other site;
(c) $f_{\textrm{H}}^2$ intermediate states on both sites.
Arrows show $k_FR_{nn}$.}
\label{FigureQuad}
\end{figure}

Fig. \ref{FigureQuad} shows the spatial
dependence 
of the quadrupole couplings $\mib{D}_{3g}$ and $\mib{D}_{5g}$
for $\mib R\parallel (001)$.
With the intermediate state $f^0$, there are only two independent coupling constants for any distance along (001).  
The four-fold symmetry around the pair is high enough to lead to this property in consistency with the argument of ref.\citen{Shiba}.
In the case of $f_{\text H}^2$ as intermediate states,  there are four different coupling constants. 
The anisotropy among them reflects the difference between  the $z$ component, which is parallel to $\mib R$, and $x$ or $y$ component perpendicular to $\mib R$.
For the nearest neighbors we obtain 
$\mib{D}_{5g}^{\alpha \alpha}(\xi_1,\xi_2;\hat{z}R_{nn})>0$ 
for all $\alpha = x,y,z$ 
and all combinations of $\xi_1$ and $\xi_2$.
This positive sign favors the simple AFQ ordering as observed in CeB$_6$, and is also consistent with the temperature dependence of the elastic constants.\cite{Goto,Nakamura,Suzuki}

On the other hand, the spatial dependence of 
$\mib{D}_{3g}$ does not depend much on the 
intermediate state $f^0$ or $f_{\text H}^2$.
At the nearest-neighbor distance $k_F R_{nn}\sim 3.09$, we obtain the anisotropic behavior: $D_{3g}^{zz}<0$ and $D_{3g}^{xx}>0$.
This result is interpreted by the property of the transfer integral;
if the 4$f$ wave functions at the site $\mib{R}_i$ and $\mib{R}_j$ are stretched toward each other, the transfer integral from $\mib{R}_i$ to $\mib{R}_j$ becomes large and the energy is lowered.  In other words, we should have a ferro-quadrupole coupling between $O_2^0 (i)$ and $O_2^0 (j)$ at short distance. 
The relation $D_{3g}^{zz} < D_{3g}^{xx}$ will be discussed from a more mathematical point of view in \S\ref{bond}.
Since the absolute magnitude of $\mib{D}_{3g}$ is similar to that of $\mib{D}_{5g}$, 
it is not possible to explain the stability of the phase II in CeB$_6$ with the $\mib{D}_{5g}$ order only from the above result.  However, the observed antiferro-type order is consistent with our simple theory for the $\mib{D}_{5g}$ interaction.

\subsection{Dipole-dipole interaction}
For convenience, we regard the vector representations $\Gamma=4u1,  4u2$ as representing dipoles, although they actually mix with octupoles under the cubic symmetry.
The pure octupoles corresponds to $\Gamma=2u, 5u$ and are studied later.
The intersite interaction between two ``dipoles'' separated by $\mib R$ is then given by 
\begin{align}
H_D &= \frac{|V_{k_F}|^4}{E_a} \sum_{\xi_1,\xi_2}
\frac{1}{E_{\textrm{exc}}(\xi_1)E_{\textrm{exc}}(\xi_2)} \nonumber \\
& \times \sum_{\alpha,\beta}^{x,y,z}
[D_{4u1}^{\alpha \beta} (\xi_1, \xi_2; \mib R)\sigma^{\alpha}(1)\sigma^{\beta}(2)
+ D_{4u2}^{\alpha \beta}(\xi_1, \xi_2; \mib R)\eta^{\alpha}(1)\eta^{\beta}(2)],
\end{align}
In addition to the nearest-neighbor interaction, we are much interested in the next-nearest-neighbor interaction in this case.  
Especially, we shall examine whether the sign of the pseudo-dipolar interaction is consistent with the one required by the phenomenological model.\cite{Kusunose}

\begin{figure}[htbp]
\begin{center}
	\includegraphics[width=8cm, clip]{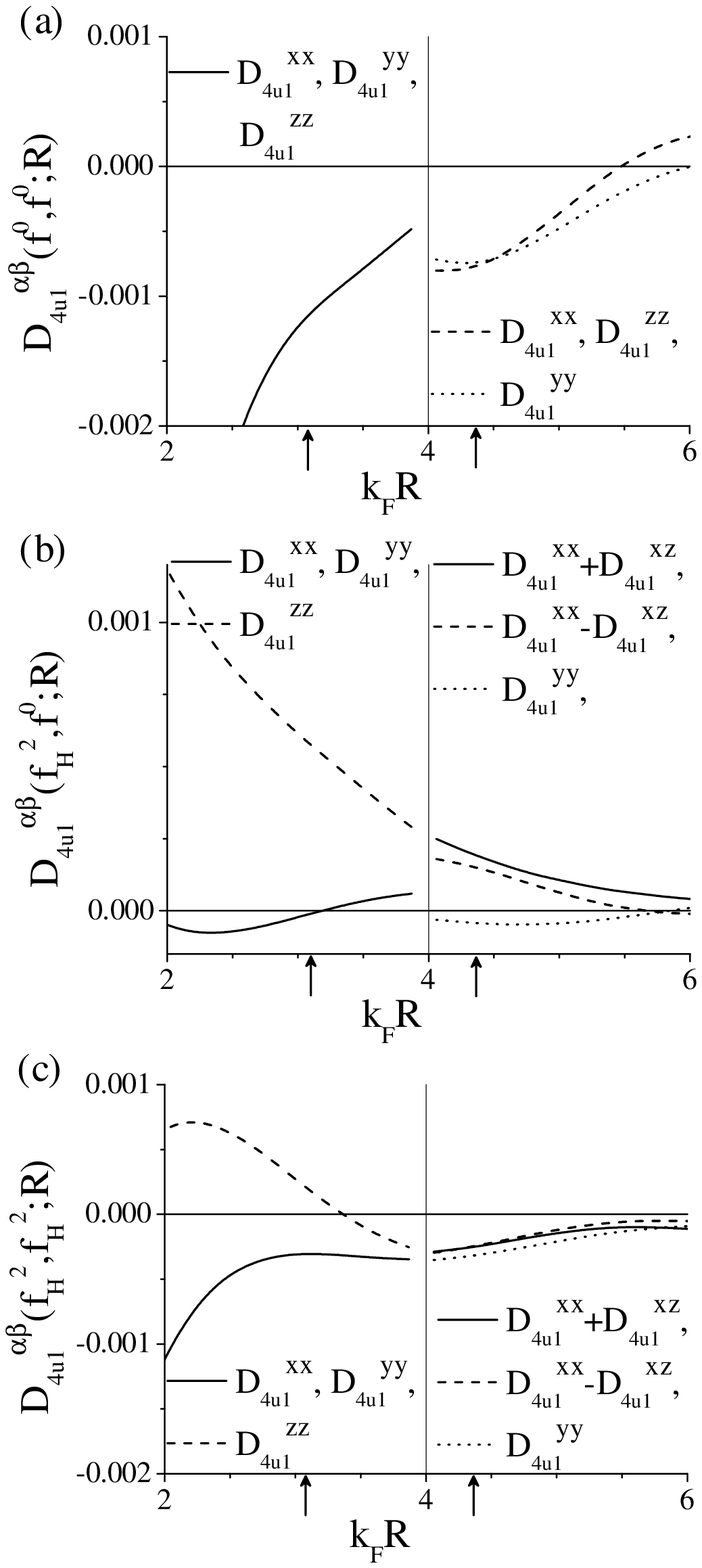}
\end{center}
\caption{The spatial dependence of dipole-dipole interaction $\mib{D}_{4u1}$.
The part with $k_FR<4$ shows the case of 
$\mib{R}=(0,0,R)$, and the other part with $k_FR>4$ gives the result 
for $\mib{R}=(R,0,R)/\sqrt{2}$.
Arrows show $k_FR_{nn}$ and $k_FR_{nnn}$.}
\label{Figure4u1}
\end{figure}

Fig.\ref{Figure4u1} shows 
the spatial dependences of $\mib{D}_{4u1}$.
The direction of $\mib R$ in the region $k_FR < 4$ is parallel to (001), and the position of a nearest-neighbor at $k_FR_{nn}\sim 3.09$ is included.
For $k_FR > 4$ we take $\mib R$ parallel to (101), which includes a next-nearest-neighbor position with $k_FR_{nnn}\sim 4.37$.
In Fig.\ref{Figure4u1}(a) we obtain isotropic 
$\mib{D}_{4u1}(f^0,f^0;\hat{z}R)$, 
which is consistent with the theory of Shiba {\it et al.} who started from a different model.
We note that the isotropy in our case is generated only after the projection to the $\Gamma_8$ CEF states.
Before projection, a finite component with $L=2$ in eq.(\ref{tensorproduct}) gives anisotropic dipole interaction.
On the other hand, the next-nearest-neighbor interaction
$\mib{D}_{4u1}(f^0,f^0;\mib R_{nnn})$ becomes anisotropic, but is still isotropic in the $xz$-plane.
We shall discuss the origin of the anisotropy in detail later, and only
remark here that the $f^0$ intermediate state leads to ferromagnetic interaction at both $\mib R_{nn}$ and $\mib R_{nnn}$.

With $f_{\rm H}^2$ included as intermediate states, the $\Gamma_{4u1}$ interaction becomes anisotropic even for $\mib R$ along (001) as seen in (b) and (c).
In contrast with the case of $\Gamma_{3g}$ where the $z$-component favors the ferro-quadrupole configuration, 
the $z$-component $D_{4u1}^{zz}$ tends to be antiferro-magnetic with $f_{\rm H}^2$ as seen in (b) and (c). 

We have checked that $\mib{D}_{4u1}(f_{\textrm H}^2,f^0;\hat{z}R)$ tends to cancel $\mib{D}_{4u1}(f^0,f^0;\hat{z}R)$ and $\mib{D}_{4u1}(f_{\textrm H}^2,f_{\textrm H}^2;\hat{z}R)$ at short distance (not shown in Figure).
This is because the $f^0$ intermediate state tends to align the spins of $f$ and conduction electrons antiparallel, 
while $f_{\text H}^2$ intermediate states tend to align them parallel.
At short distance the propagation of conduction electrons just connects this tendency on two sites.

\begin{figure}[htbp]
\begin{center}
	\includegraphics[width=8cm, clip]{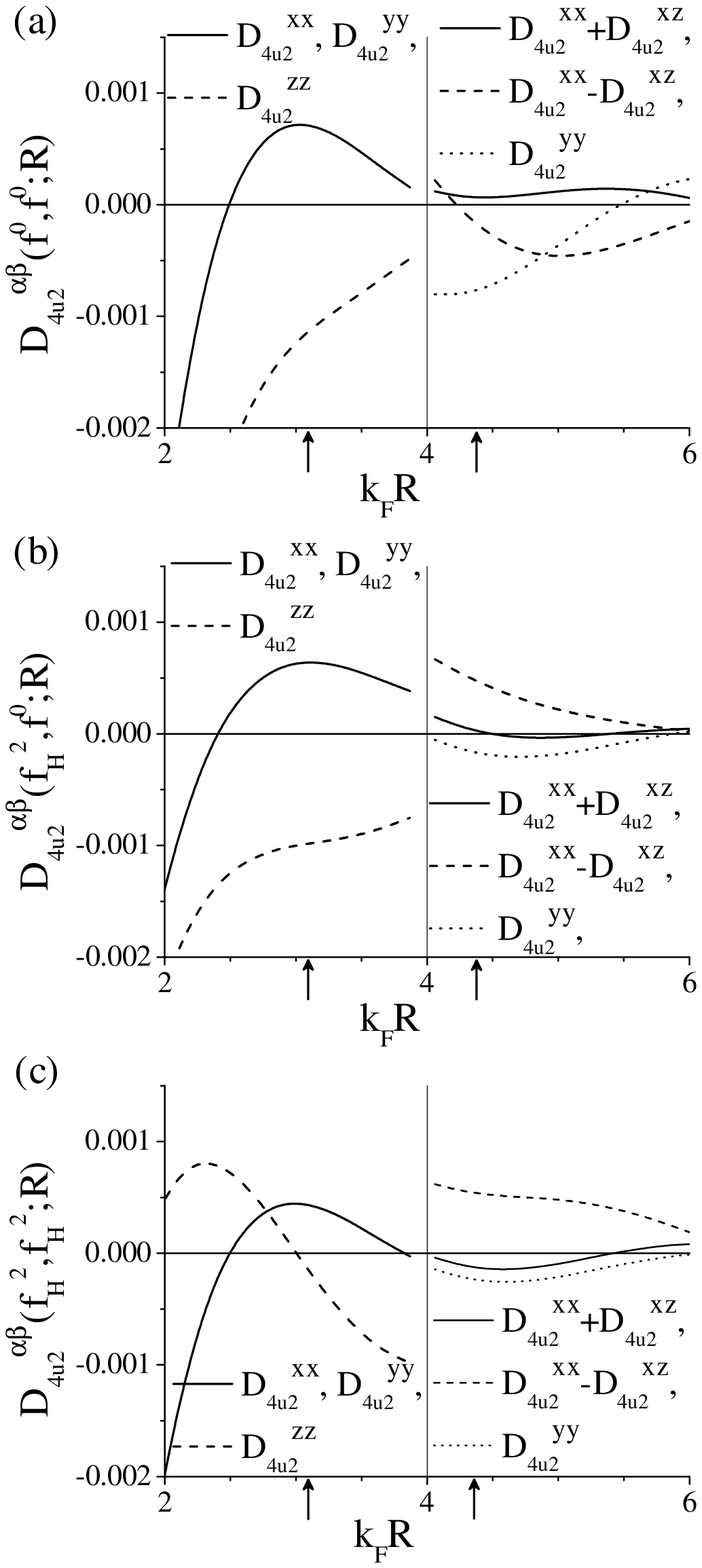}
\end{center}
\caption{The spatial dependence of dipole-dipole interaction $\mib{D}_{4u2}$.
The cases (a), (b), (c) are the same as in Fig.\ref{FigureQuad}.
And the defintion of $\mib{R}$ are the same as in Fig.\ref{Figure4u1}.
Arrows show $k_FR_{nn}$ and $k_FR_{nnn}$.}
\label{Figure4u2}
\end{figure}

Fig.\ref{Figure4u2} shows the spatial dependence of $\mib{D}_{4u2}$ for $\mib R$ along (001) and (101).
There appear off-diagonal elements $D_{4u2}^{xz} = D_{4u2}^{zx}$ for $\mib{R}$ along (101) direction.
By symmetry we obtain $D_{4u2}^{xx} = D_{4u2}^{zz}$ and $D_{4u2}^{xz} = D_{4u2}^{zx}$.  Then it is convenient to take the principal axes as 
$\hat{e}_\parallel=(\hat{x}+\hat{z})/\sqrt 2$, 
$\hat{e}_\perp=(\hat{x}-\hat{z})/\sqrt 2$ and 
$\hat{y}$.
Along the pair axis $\hat{e}_\parallel$ 
the relevant eigenvalue is given by $D_{4u2}^{xx}+D_{4u2}^{xz}$,
and along $\hat{e}_\perp$ by $D_{4u2}^{xx}-D_{4u2}^{xz}$.
The interaction is anisotropic even along (001).  This can be understood if one recalls that $\Gamma_{4u2}$ involves the orbital flip, in contrast with 
$\Gamma_{4u1}$.  Hence the anisotropy 
occurs more likely than in $\Gamma_{4u1}$ along (001).  
At the nearest-neighbor distance, the anisotropy is opposite to that of $\Gamma_{4u1}$.  
We expect that the sum reduces the anisotropy of each contribution at $\mib R_{nn}$.
For the next-nearest-neighbor interaction, the anisotropy in Fig.\ref{Figure4u2}(a) with $f^0$ intermediate state is opposite from that in (b) and (c).
In (b) and (c),  $D_{4u2}^{xx} -D_{4u2}^{xz}$ shows antiferro-magnetic coupling and is larger than the other components in absolute value.
This situation corresponds to $K(\mib R_{12})<0$ in eq.(\ref{pseudo-dipole}) and is consistent with the phenomenological theory for the phase III of CeB$_6$.\cite{Kusunose}
However, the result in (a) cannot be interpreted in terms of eq.(\ref{pseudo-dipole}), since they differ significantly in all three principal axes.
As for the behaviour at short range, which is not seen in Fig.\ref{Figure4u2}, the intermediate states $f^0$ and $f_{\text H}^2$ give additive contribution to $\mib{D}_{4u2}$.
This is because the on-site 
orbital exchange and the simultaneous exchange of orbit and spin, which are partly represented by $\mib{D}_{4u2}$, have the same sign for $f^0$ and $f_{\text H}^2$ intermediate states.

\subsection{Octupole-octupole interaction}
\begin{figure}[htbp]
\begin{center}
	\includegraphics[width=8cm, clip]{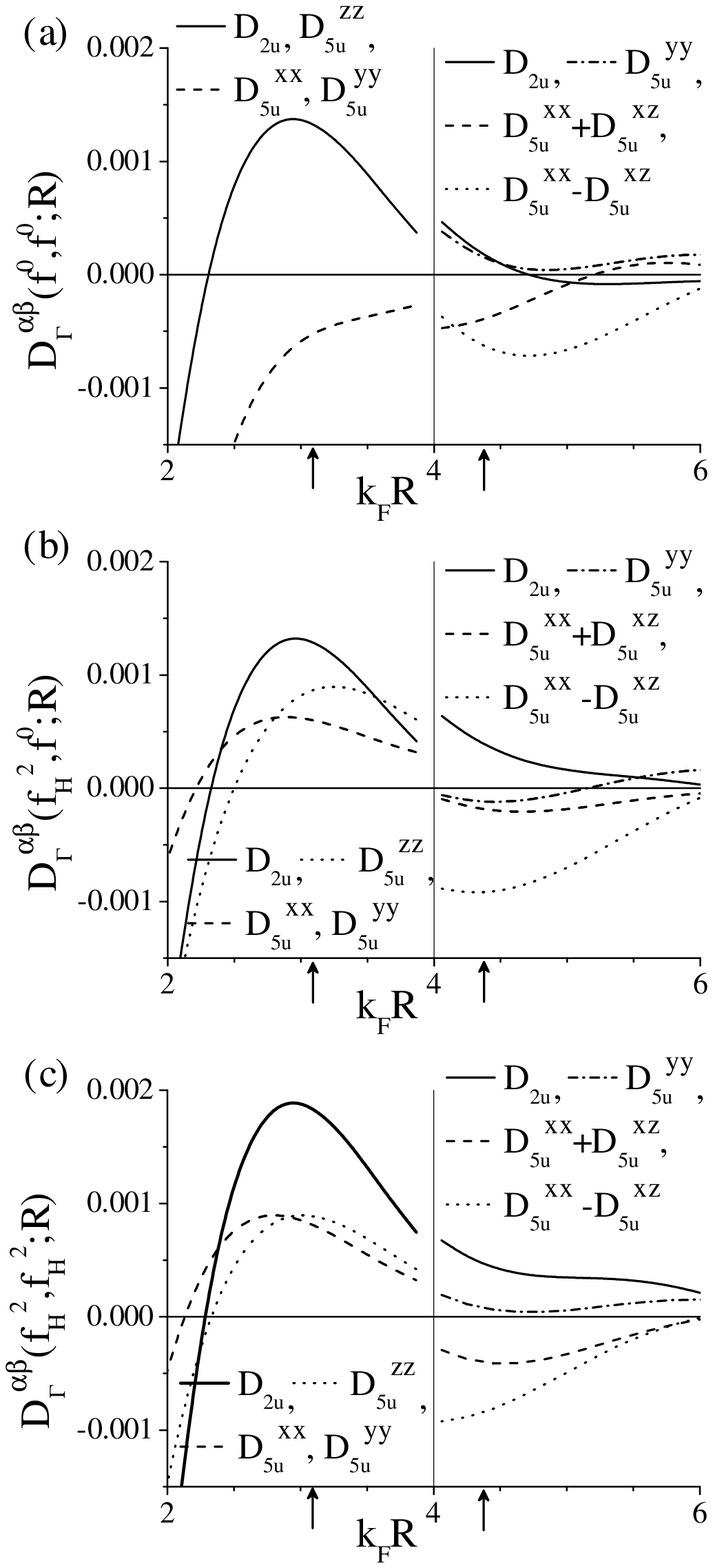}
\end{center}
\caption{The spatial dependence of octupole-octupole interactions $D_{2u}$ and $\mib{D}_{5u}$.
The cases (a), (b), (c) are the same as in Fig.\ref{FigureQuad}.
And the definition of $\mib{R}$ are the same as in Fig.\ref{Figure4u1}.
Arrows show $k_FR_{nn}$ and $k_FR_{nnn}$.}
\label{FigureOct}
\end{figure}

We now turn to the octupolar interaction.  
This interaction plays an important role in understanding the NMR and neutron scattering results in CeB$_6$ consistently,\cite{Sakai} and also the phase diagram in magnetic fields.\cite{Ohkawa,Shiina}
The interaction is given by
\begin{align}
H_O &= \frac{|V_{k_F}|^4}{E_a} \sum_{\xi_1,\xi_2}
\frac{1}{E_{\textrm{exc}}(\xi_1)E_{\textrm{exc}}(\xi_2)} \nonumber \\
&\times [
D_{2u}(\xi_1, \xi_2; \mib R) \tau^y (1)\tau^y (2)
+\sum_{\alpha\beta}^{x,y,z}D_{5u}^{\alpha \beta} (\xi_1, \xi_2; \mib R) \zeta^\alpha (1)\zeta^\beta (2)
].
\end{align}
Fig.\ref{FigureOct} shows the spatial dependences of the octupole-octupole interactions $D_{2u}$
 and $\mib{D}_{5u}$ for $\mib{R}$ along (001) and (101). 
At the nearest-neighbor distance, we obtain $D_{2u}>0$ irrespective of the intermediate states, which means the anti-ferro octupole coupling. 
On the other hand, 
$\mib{D}_{5u}(f^0,f^0;\mib R)$ has a different spatial dependence from 
that of $\mib{D}_{5u}(f_{\textrm H}^2,f^0;\mib R)$ or $\mib{D}_{5u}(f_{\textrm H}^2,f_{\textrm H}^2;\mib R)$.
At the next-nearest-neighbor distance, 
$D_{5u}^{xx}\!-\!D_{5u}^{xz}$ is negative and is larger than the other components in absolute value for cases (a), (b) and (c).
This means that $\Gamma_{5u}$ moments on next-nearest-neighbor pair tend to be parallel to each other and along $\hat{e}_\perp$, namely perpendicular to $\mib{R}$.
This feature of "anti-pseudo-dipole" interaction is also consistent with the phenomenological theory for 
the transition from III to III' under magnetic field.\cite{Kusunose}

The above results for various kinds of multipolar interactions
show that the dipolar, quadrupolar and octupolar interactions have the same order of magnitude.
Especially the octupolar interaction between $\Gamma_{2u}$ type moments is as strong as the quadrupolar interaction between $\Gamma_{5g}$ type moments.
If we consider only $f^0$, the results are consistent with the previous symmetry analysis.\cite{Shiba}
In addition, the $\Gamma_{5u}$-type octupolar interactions 
show the range dependence similar to that of the 
$\Gamma_{5g}$-type quadrupolar interactions 
for a given set of intermediate states.

\subsection{Relation to the bond density}\label{bond}
In order to understand better the nature of the intersite interaction,
we apply the 
the idea of ``bond density" of Shiba {\it et al.} to the present model.\cite{Shiba}
We can 
trace difference of spatial dependences, and identify the independent components describing the interaction. 
For simplicity we consider only the 4$f^0$ intermediate state.
The bond density is defined by
\begin{align}
&b(M,M',k,\mib{R}) \nonumber \\
&=\sum_{\sigma}
\langle j_1 M |3 M\!-\!\sigma, \frac{1}{2}\sigma \rangle
\langle 3 M'\!-\!\sigma, \frac{1}{2}\sigma |j_1 M' \rangle 
\int d\Omega_{\mib{k}} Y_{3 M-\sigma}^\ast(\Omega_{\mib{k}})Y_{3 M'-\sigma}(\Omega_{\mib{k}})
e^{i \mib{k}\cdot \mib{R}} \nonumber \\
&\equiv
\sum_{l=0}^\infty b_l(M,M',\hat{R}) j_l(kR).
\label{bond-M}
\end{align}
In the above we have used eq.(\ref{dOmegaK}) to decompose 
the bond density into partial-wave components $b_l$.
Each component $b_l$ is projected to the $\Gamma_8$ subspace as
\begin{equation}
B_l(\nu \sigma, \nu' \sigma',\hat{R}) = \sum_{M,M'}
\langle \nu \sigma| M\rangle b_l(M,M',\hat{R})\langle M'|\nu' \sigma' \rangle.
\label{bond-nu}
\end{equation}

For deriving the spatial dependence and the strength of each component, 
we take as before the quantization axis $z$ along the pair of Ce ions, and take the unit vector $\hat{z}$.
We obtain $b_l(M,M',\hat{z})\propto \delta_{MM'}$, 
since eq.(\ref{dOmegaK}) requires $M'-M=m$ and 
$Y_{lm}(\hat{z})$ vanishes unless $m=0$.
This in turn gives the diagonal property: 
$B_l(\nu \sigma, \nu' \sigma',\hat{z})\propto \delta_{\nu\nu'}\delta_{\sigma\sigma'}$.
The projected interaction between a pair along the $z$ axis is then given by
\begin{equation}
D^{(\nu,\nu')}(\sigma,\sigma', \hat{z}R) = \frac{(k_Fa)^4}{8\pi^4}\sum_{l,l'}
	B_l(\nu\sigma,\nu\sigma,\hat{z})
	B_{l'}(\nu'\sigma',\nu'\sigma',\hat{z})
	Q_{ll'}(k_FR).
\label{projected-int}
\end{equation}
For the nearest-neighbor interaction in the cubic lattice, the above choice of the quantization axis is consistent with the CEF states.
On the other hand, the next-nearest-neighbor interaction in the $\Gamma_8$ subspace requires the change of the quantization axis since the orbital index $\nu$ is given for the cubic axis.\cite{Shiba}

As was shown 
in ref,\cite{Shiba} the high-symmetry around the $z$ axis makes $D^{(\nu,\nu')}(\sigma,\sigma', \hat{z}R)$ independent of $\sigma$, i.e., 
$D^{(\nu,\nu')}(\sigma,\sigma', \hat{z}R)= D^{(\nu,\nu')}(\hat{z}R)$ .
For any representation $\Gamma$, we can represent the relevant interaction matrix
$\mib{D}_\Gamma (f^0,f^0;\hat{z}R)$ in terms of  $D^{(\nu,\nu')}$.
For example, components the of quadrupole-quadrupole interaction $\mib{D}_{3g}$ 
is given by 
\begin{align}
D_{3g}^{zz}(f^0,f^0;\hat{z}R) =& [D^{(+,+)}(\hat{z}R)+D^{(-,-)}(\hat{z}R)]/8,  \label{BondD3g}\\
D_{3g}^{xx}(f^0,f^0;\hat{z}R) =& [D^{(+,-)}(\hat{z}R)+D^{(-,+)}(\hat{z}R)]/8. 
\end{align}
Components 
$D_{5g}^{\alpha \alpha}(f^0,f^0;\hat{z}R)$ with $\alpha = x,y,z$ 
turns out to be the same as $D_{3g}^{xx}(f^0,f^0;\hat{z}R)$ in consistency with the results in Fig.\ref{FigureQuad}.

The sign of the anisotropy is clearly seen by 
retaining $l,l'$ up to two in the partial wave decomposition.
This approximation is reasonable because the anisotropic range functions $|Q_{ll'}(x)|$ with $l,l'\neq 0$ are smaller than $|Q_{00}(x)|$ at small $|x|$.
We then obtain the components of the $\Gamma_3$ 
quadrupole-quadrupole interaction as follows:
\begin{align}
D_{3g}^{zz}(f^0,f^0;\hat{z}R) &\simeq \frac{(k_Fa)^4}{32\pi^4}
\left[Q_{00}(k_FR)
+\left(\frac{8}{7}\right)^2 Q_{22}(k_FR)\right], \\
D_{3g}^{x x}(f^0,f^0;\hat{z}R) &\simeq \frac{(k_Fa)^4}{32\pi^4}
\left[Q_{00}(k_FR)
-\left(\frac{8}{7}\right)^2 Q_{22}(k_FR)\right].
\label{xx-vs-zz}
\end{align}
In general the hybridization of conduction electron causes ferro-type coupling between ions at short distance, and we have indeed $Q_{22}(x)<0$ at $x$ around 3.
Then for the quadrupole-quadrupole interactions we have a relation 
$D_{3g}^{zz} < D_{3g}^{x x}$ at the nearest-neighbor distance.
This is indeed satisfied by the results shown in Fig.\ref{FigureQuad}.

Table~\ref{PWexpand} shows interactions between operators within $\Gamma$ 
by the coefficients of $D^{(\nu,\nu')}$ listed in the top column.
Components $D_{\Gamma}^{\alpha \alpha}(f^0,f^0;\hat{z}R)$ with $\alpha = x,y$ are shown with $\perp$, and $D_{\Gamma}^{zz}(f^0,f^0;\hat{z}R)$ with $\parallel$.
From Table~\ref{PWexpand},  it turns out that $D_{4u2}^\parallel <D_{4u2}^\perp$, 
and
$D_{5u}^\perp < D_{5u}^\parallel$.
These two relations are consistent with the results in Figs \ref{Figure4u2} and \ref{FigureOct}.

\begin{table}
\caption{The expansion coefficients of interactions $\mib{D}_{\Gamma}(f^0,f^0,\hat{z}R)$ by $D^{(\nu,\nu')}(\hat{z}R)$.
}
\label{PWexpand}
	\begin{center}
	\begin{tabular}{cccc} \hline
	$\Gamma$ &component& $D^{(+,+)}\!+\!D^{(-,-)}$ & $D^{(+,-)}\!+\!D^{(-,+)}$ \\ \hline
	4u1 & $\perp,\parallel$		& 1/8	& 0 	\\
	4u2 & $\perp$		& 1/32	& 3/32 	\\
	\,  & $\parallel$	& 1/8	& 0 	\\ \hline
	3g  & $\perp$		& 0		& 1/8 	\\
	\,  & $\parallel$	& 1/8	& 0		\\
	5g  & $\perp,\parallel$		& 0		& 1/8 	\\ \hline
	2u  & $ $			& 0		& 1/8	\\
	5u  & $\perp$		& 3/32	& 1/32 	\\
	\,  & $\parallel$	& 0 	& 1/8 	\\ \hline
	\end{tabular}
	\end{center}
\end{table}
\subsection{Anisotropy of the $\mib\sigma$-$\mib\sigma$ Interaction}

It has been found that the exchange interaction between a pair of pseudo-spins $\mib\sigma$ becomes isotropic in the case where the intermediate states of hybridization is either $4f^0$ or $4f^{14}$.\cite{Shiba,Shiba-Yb4As3}
The isotropy follows 
independently of the local symmetry around the pair.
As described in the main text, however, the next-nearest-neighbor exchange interaction in the $\Gamma_8$ subspace is anisotropic. 
Here we clarify the condition under which the effective exchange becomes isotropic or anisotropic.

\subsubsection{No orbital degeneracy}

Let the CEF ground state be $\Gamma$ which has the Kramers degeneracy, but no orbital one.  Within this CEF level,  the on-site exchange interaction at a $4f^1$ site $\mib R_i$
can be written in the form
\begin{equation}
H_{\rm exch} (i) = J_\Gamma \sum_{\sigma\sigma'} 
X_{\sigma\sigma'}^\Gamma (i) 
c_{\Gamma\sigma'}^\dagger (i) c_{\Gamma\sigma} (i),
\label{H-Gamma}
\end{equation}
where 
$X_{\sigma \sigma'}^\Gamma (i) $ denotes either the transition from $\sigma'$ to $\sigma$ within the Kramers doublet, or the projection to $\sigma$ in the case of  $\sigma= \sigma'$. 
The conduction states are given in terms of the Wannier states.  
We note that this form of the exchange results only if the intermediate state is isotropic.
The intersite interaction between $\mib R_i$ and $\mib R_j$ is determined by 
the average:
\begin{equation}
\langle 
c_{\Gamma\sigma}^\dagger (i) 
c_{\Gamma\sigma'}(j)
\rangle,
\end{equation}
and its Hermitian conjugate.
With the time reversal symmetry preserved in the paramagnetic state
the average is nonzero only if $\sigma=\sigma'$.
Then the intersite interaction is proportional to the permutation operator
\begin{equation}
P_{ij}= \sum_{\sigma\sigma'} X_{\sigma\sigma'}^\Gamma (i) 
X_{\sigma'\sigma}^\Gamma (j).
\end{equation}
As in the main text we introduce the pseudo-spin operator  
$ 
\sigma_x (i) = 
X_{\uparrow\downarrow}^\Gamma (i) + 
X_{\downarrow\uparrow}^\Gamma (i) 
$
and other components $\sigma_y (i)$ and $\sigma_z (i)$ in an analogous way.
Here the first component of the Kramers pair is indicated by $\uparrow$ and the second one by $\downarrow$.
Because of the identity
\begin{equation}
P_{ij} = \frac 12 [
\mib\sigma (i) \cdot \mib\sigma (j) +1
],
\end{equation}
the effective exchange is found to be isotropic.
We emphasize that the isotropic exchange is independent of the spatial symmetry around the pair, since it is the time-reversal symmetry 
that protects the isotropy.

\subsubsection{Orbital degeneracy}
Now we turn to the case of orbital degeneracy such as $\Gamma_8$.   
The exchange interaction 
now has orbital components specified by $a,b$.  
Provided the intermediate state of hybridization is isotropic, only  the permutation of $\Gamma_8$ components happens between the conduction states.  Namely we obtain
\begin{equation}
H_{\rm exch} (i) = J_\Gamma \sum_{\sigma\sigma' ab} 
X_{a\sigma\ b\sigma'}^{\Gamma}(i) 
c_{\Gamma b\sigma'}^\dagger (i) c_{\Gamma a\sigma} (i),
\label{Hexch-Gamma}
\end{equation}
in the Wannier representation for conduction states.
We now examine the possibility
\begin{equation}
\langle 
c_{\Gamma a\uparrow}^\dagger (i) 
c_{\Gamma b\downarrow}(j)\rangle \neq 0,
\label{spin-off}
\end{equation}
with $a\neq b$.  
In the case of the nearest-neighbor pair, the four-fold symmetry around the pair excludes the possibility, as discussed in \S 3.4. 
Then the $\Gamma_{4u1}$ intersite interaction becomes isotropic in the spin space.  This situation was already noticed in ref.\citen{Shiba}.

However, if the second site measured from the first has the direction $\hat{R}$ other than the four-fold axis, we may have
$$ 
Y_{lm}(\hat{R})\neq 0
$$
with $m=\pm 1$ and $\pm 3$ for $|m| \leq l\leq 6$ in eq.(\ref{dOmegaK}).\cite{UpperboundL}
This is indeed the case of a next-nearest-neighbor pair where the first site is at the origin and the second one is at $(a,0,a)$.
From eqs.(\ref{dOmegaK}) and (\ref{JMJM}), it is seen that 
a conduction electron including the component $j_z=-1/2$  from the first site can propagate to another state with $j_z=5/2$ in the second site with $m=-3$,
and to $j_z=-3/2$ with $m=1$.
This situation realizes the possibility eq.(\ref{spin-off}).
Note that eq.(\ref{spin-off}) does not necessarily mean the breakdown of the time reversal.  
The operators $\tau_y\sigma_x$ and $\tau_y\sigma_y$, which are even under time reversal, describe the simultaneous flips of two kinds of quasi spins; orbital and Kramers pairs. 

Let us analyze a part of the exchange interaction which involves only the pseudo-spin component $\mib\sigma$, but not the other component $\mib\tau$.   In the cubic symmetry this corresponds to the representation $\Gamma_{4u1}$. 
The absence of the $\mib\tau$ component means that on-site process is diagonal with respect to the orbital index.
However the Kramers index may have such combination with $\bar{a} \neq a$ as 
\begin{equation}
\sum_{a}
X_{a\uparrow a\downarrow}^{\Gamma} (i) 
X_{\bar{a}\uparrow \bar{a}\downarrow}^{\Gamma} (j) = 
\sum_{\pm}\frac{1}{16}  
[\sigma^x(i) +i\sigma^y(i)] [1\pm \tau^z(i)] 
[\sigma^x(j) +i\sigma^y(j)] [1\mp \tau^z(j)] ,
\end{equation}
and its Hermitian conjugate.  
Expansion of the above form gives rise to an anisotropic term:
\begin{equation}
\sigma^x(i) \cdot\sigma^x(j) - \sigma^y(i) \cdot\sigma^y(j),
\end{equation}
in the pseudo-spin space $\mib\sigma$.
We understand in this way the results in Fig.\ref{Figure4u1} which indeed show anisotropy in the pseudo-spin $\mib\sigma$.

We now discuss why there is no off-diagonal component in Fig.\ref{Figure4u1}(a).
It is convenient to use the bond-density defined by eqs.(\ref{bond-M}) and (\ref{bond-nu}) with the properties
\begin{align}
&b_l(M,M',\hat{R}) \propto (-1)^{j_1-M'}
\sum_m
\begin{pmatrix}
	j_1 & l & j_1 \\
	M & m & -M'
\end{pmatrix}
Y_{lm}(\hat{R}), \label{bprop} \\
&b_l(-M',-M,\hat{R}) = b_{l}^{\ast}(M,M',\hat{R}),\\
&B_l(\nu \bar{\sigma}, \nu' \bar{\sigma}', \hat{R}) = B_{l}^{\ast}(\nu' \sigma', \nu \sigma, \hat{R}), 
\label{B-rev}
\end{align}
with $\bar{\sigma}=-\sigma$.
Then $D_{4u1}^{zx}(f^0,f^0,\mib{R})$ is proportional to
\begin{equation}
\sum_{l,l'}Q_{ll'}(k_FR) 
\sum_{\nu, \nu'}
\sum_{\alpha\beta\gamma\delta}
(\sigma^z)_{\alpha\beta}
(\sigma^x)_{\gamma\delta}
 B_l(\nu\alpha, \nu'\delta,\hat{R})B_{l'}(\nu'\gamma, \nu\beta, \hat{R}).
\label{ZXcomp}
\end{equation}
Since $\hat{R}$ is in the $xz$-plane, 
$Y_{lm}(\hat{R})$ is real with the azimuthal angle $\phi=0$.
Then using eq.(\ref{B-rev}) we see that $B_l$ is also real, 
and hence eq.(\ref{ZXcomp}) is zero with $Q_{ll'}(k_FR)=Q_{l'l}(k_FR)$.
Therefore $D_{4u1}^{zx}(f^0,f^0,\mib{R})$ vanishes when $\hat{R}$ is in the $xz$-plane.

\section{Summary}

We have derived the multipolar interaction between Ce ions from the Anderson-type model with orbital degeneracy using the free-electron model for the conduction electrons, 
with attention to the electronic structures of the phases II and III in CeB$_6$.
As for intermediate states, we have considered not only 4$f^0$ but 4$f^2$ Hund's-rule ground states,
and have shown that the anisotropic exchanges which are expressible in terms of pseudo-spin operators arise.
From the effective Hamiltonian, we have calculated coupling strengths for dipole-dipole, quadrupole-quadrupole, and octupole-octupole interactions.
In this paper we have assumed the single spherical Fermi surface, which is evidently oversimplified if we compare with the experimental situation in CeB$_6$.
Moreover there should be the Coulombic exchange in addition to the hybridization as another source of the intersite interaction.
In spite of our simplification, the derived coupling constants give fairly 
consistent account of the AFQ order of $\Gamma_{5g}$ type in the phase II.
The pseudo-dipole coupling constant of $\Gamma_{4u2}$ type and the 
anti-pseudo-dipole coupling constant of $\Gamma_{5u}$ type are both in consistency with the phenomenological model,\cite{Kusunose}
provided 4$f_{\text H}^2$ intermediate states are taken into account.

In relation to the previous work,\cite{Shiba} we calculated the explicit spatial dependence of multipolar couplings within free-electron approximation.
The all coupling constants of multipolar interaction in this system have the same order of magnitude within this simplified model.
Especially in considering only 4$f^0$ intermediate state, our calculation is consistent with the previous symmetry analysis on the couplings between nearest neighbors. 
The octupole-octupole interaction between $\Gamma_{2u}$ type moments is as strong as the quadrupole-quadrupole interaction between $\Gamma_{5g}$ type moments.
For next-nearest-neighbors, multipolar couplings become anisotropic even if only 4$f^0$ intermediate state is considered.

As for the 4$f^2$ intermediate states, we have considered only the 
Hund's-rule ground states.
Inclusion of other states in the 4$f^2$ multiplets should have a tendency to enhance the simultaneous exchange of spin and orbit.
This can be seen from the results in Table~\ref{ConsequeceOfIntermediate}.
Namely, simultaneous spin and orbital exchange has additive contribution from both intermediate states with $S=0$ and $S=1$.
After projection to the $\Gamma_8$ states, the orbital exchange is described by the pseudo-spin $\mib\tau$, and the spin exchange by $\mib\sigma$.  Although the spin-orbit interaction mixes the original orbital and spin quantum numbers considerably, the sign of the on-site exchange is mainly determined by the antisymmetry of $4f^2$ states, and is robust against such mixing.
After projection to the $\Gamma_8$ states, the simultaneous spin-orbital exchange takes the form of  
$\Gamma_{5g}$, $\Gamma_{4u2}$ and $\Gamma_{5u}$ on-site interactions.  
It is suggestive that these interactions seem relevant to realizing the phase diagram of actual CeB$_6$.\cite{Kusunose}

\section*{Acknowledgement}
The authors are grateful to Prof. T. Takahashi for useful information on 4$f$ levels in CeB$_6$.
And one of us (G. S.) would like to thank H. Kusunose and K. Kubo for helpful suggestions.
\appendix

\section{Integrals involving spherical Bessel functions}
In this appendix we derive analytic expression of the range functions.
We need to calculate integrals such as
\begin{equation}
q(l_1, l_2; z) \equiv \int_0^z x^2 \, dx
\{
j_{l_1}(x)\phi_{l_2}(x) + j_{l_2}(x)\phi_{l_1}(x)
\}, \label{eq-Ab-00}
\end{equation}
where $l_1$ and $l_2$ are even integers, and
$\phi_l(x)$ is defined by
\begin{eqnarray}
\phi_l(x) \equiv \int_0^\infty \frac{{x'}^2}{x^2 - {x'}^2} j_l(x')dx'. \label{eq-Ab-01}
\end{eqnarray}
The function 
$Q_{l_1 l_2}(k_F R)$ used in the main text
is related to $q(l_1,l_2;k_FR)$ by
\begin{equation}
Q_{l_1 l_2}(k_F R) = (k_F R)^{-4}q(l_1,l_2;k_F R).
\end{equation}
Larsen has already calculated $Q_{l_1 l_2}(x)$ in the case of $l_1 = l_2$,\cite{Larsen}
but we need $Q_{l_1 l_2}(x)$ also with $l_1 \neq l_2$. 

Using the result in ref.\citen{Larsen}, $\phi_l(x)$ is given by
\begin{equation}
\phi_l(x) = \frac{\pi}{2} \left[ x n_l(x) + \sum_{m=2,4,\ldots}^{m=l}\frac{(l+m-1)!!}{(l-m)!!}\frac{1}{x^m} \right], \label{eq-Ab-24}
\end{equation}
where $n_l(x)$ is the spherical Neumann function of order $l$.
Spherical Bessel function of order $l$ can be written by using trigonometric functions:
\begin{equation}
j_l(x) = C_l(x)\cos x + S_l(x)\sin x, \label{SBessel}
\end{equation}
where $C_l(x)$ and $S_l(x)$ are the polynomial expressions of $x^{-1}$,
while spherical Neumann function of the same order is given by
\begin{equation}
n_l(x) = C_l(x)\sin x - S_l(x)\cos x. \label{SNeumann}
\end{equation}
We obtain the analytic expression for $q(l_1,l_2,z)$ by using eqs.(\ref{eq-Ab-24}), (\ref{SBessel}) and (\ref{SNeumann}) and by performing the following partial integrals successively:
\begin{align}
\int dx \frac{\sin x}{x^p} &= -\frac{\sin x}{(p-1)x^{p-1}}+\frac{1}{p-1}\int dx \frac{\cos x}{x^{p-1}} ,\\
\int dx \frac{\cos x}{x^p} &= -\frac{\cos x}{(p-1)x^{p-1}}-\frac{1}{p-1}\int dx \frac{\sin x}{x^{p-1}}.
\end{align}
The final results are tabulated in Table~\ref{TableQ}.
The functions in Table~\ref{TableQ} are given in terms of the coefficients of the trigonometric functions and their related functions listed in the top row.

Among the various range functions,
$Q_{00}(k_F R)$ gives the classical RKKY interaction,\cite{Ruderman}
\begin{equation}
Q_{00}(k_F R) = 2\pi F(2k_F R),
\end{equation}
where
\begin{equation}
F(x) = (x \cos x - \sin x)/x^4.
\end{equation}
From Table~\ref{TableQ}, the asymptotic form of $Q_{l_1 l_2}(k_F R)$ for $k_F R \rightarrow \infty$ is obtained as:\cite{Coqblin}
\begin{equation}
Q_{l_1 l_2}(k_F R) \simeq (-1)^{(l_1+l_2)/2}Q_{0 0}(k_F R).
\end{equation}

\begin{table}
\caption{Dimensionless range functions $2q(l_1, l_2; x)/\pi$. The functions are given in terms of trigonometric functions, Si$(x)$, and Si$(2x)$ $\left(\text{Si}(x) \equiv \int_0^x dt \sin t/t\right)$.} \label{TableQ}
\begin{center}
\scriptsize
	\addtocounter{table}{-1}
	\caption{\!\!\!\!\!a \, The coefficient of $x^{-n}\cos x$.}
	\begin{tabular}{llrrrrrr}				\hline
		$l_1$ & $l_2$ &
		$\!x^{-9}\cos(x)$ & $\!x^{-7}\cos(x)$ & $\!x^{-5}\cos(x)$ & $\!x^{-3}\cos(x)$ & $\!x^{-1}\cos(x)$ \\ \hline
		0 & 0 &
		\, & \, & \, & \, & \, \\
		0 & 2 &
		\, & \, & \, & \, & \, \\
		0 & 4 &
		\, & \, & \, & \, & -105/2 \\
		0 & 6 &
		\, & \, & \, & -3465/4 & 1575/8 \\ \hline
		2 & 2 &
		\, & \, & \, & \, & 9 \\
		2 & 4 &
		\, & \, & \, & 315/2 & 45/2 \\
		2 & 6 &
		\, & \, & 10395 & 2205/8 & -1071/8 \\ \hline
		4 & 4 &
		\, & \, & 3675 & 175 & -5 \\
		4 & 6 &
		\, & 1091475/4 & -23625/8 & -11025/8 & 315/8 \\
		6 & 6 &
		21611205 & -4584195/4 & -255339/4 & -7119/2 & 189/4 \\
		\hline
	\end{tabular}
	
	\addtocounter{table}{-1}
	\caption{\!\!\!\!\!b \, The coefficient of $x^{-n}\sin x$, Si$(x)$ and Si$(2x)$.}
	\begin{tabular}{llrrrrrrr} \hline
	$l_1$ & $l_2$ &
	$\!x^{-10}\sin(x)$ & $\!x^{-8}\sin(x)$ & $\!x^{-6}\sin(x)$ & $\!x^{-4}\sin(x)$ &
	$\!x^{-2}\sin(x)$ & Si$(x)$	& Si$(2x)$ \\ \hline
	0 & 0 & \, & \, & \, & \, & \, & \,& \, \\
	0 & 2 & \, & \, & \, & \, & \, & 3 & -3 \\
	0 & 4 & \, & \, & \, & \, & -105/2 & -45 & 45  \\
	0 & 6 & \, & \, & \, & -10395/4 & 1575/8 & 210 & -210  \\ \hline
	2 & 2 & \, & \, & \, & \, & -9 & 3 & -3  \\
	2 & 4 & \, & \, & \, & -315/2 & 30 & 18 & -18  \\
	2 & 6 & \, & \, & -10395 & 25515/8 & -945/4 & -150 & 150 \\ \hline
	4 & 4 & \, & \, & -3675 & 1050 & 145 & 10 & -10 \\
	4 & 6 & \, & -1091475/4 & 751275/8 & 12915/2 & 105/8 & 45 & -45 \\
	6 & 6 & -21611205 & 33399135/4 & 324135/2 & 5103/2 & -504 & 21 & -21 \\ \hline
	\end{tabular}
	
	\addtocounter{table}{-1}
	\caption{\!\!\!\!\!c \, The coefficient of $x^{-n}\cos (2x)$.}
	\begin{tabular}{llrrrrrr} \hline
	$\!l_1$ & $\!l_2$ &
	$\!x^{-9}\cos(2x)$ & $\!x^{-7}\cos(2x)$ & $\!x^{-5}\cos(2x)$ &
	$\!x^{-3}\cos(2x)$ & $\!x^{-1}\cos(2x)$ & $\!x\cos(2x)$ \\ \hline
	0 & 0 &
	\, & \, & \, & \, & \, & 1/2 \\
	0 & 2 &
	\, & \, & \, & \, & \, &-1/2 \\
	0 & 4 &
	\, & \, & \, & \, & \, & 1/2 \\
	0 & 6 &
	\, & \, & \, & -3465/2 & \, & -1/2 \\ \hline
	2 & 2 &
	\, & \, & \, & \, & -9 & 1/2 \\
	2 & 4 &
	\, & \, & \, & -315/2 & 30 & -1/2 \\
	2 & 6 &
	\, & \, & -10395 & 7245/2 & -63 & 1/2 \\ \hline
	4 & 4 &
	\, & \, & -3675 & 3325/2 & -100 & 1/2 \\
	4 & 6 &
	\, & -1091475/4 & 278775/2 & -24885/2 & 210 & -1/2 \\
	6 & 6 &
	-21611205 & 47806605/4 & -2819313/2 & 95571/2 & -441 & 1/2 \\ \hline
	\end{tabular}
	
	\addtocounter{table}{-1}
	\caption{\!\!\!\!\!d \, The coefficient of $x^{-n}\sin (2x)$.}
	\begin{tabular}{llrrrrrr} \hline
	$\!l_1$ & $\!l_2$ &
	$\!x^{-10}\sin(2x)$ & $\!x^{-8}\sin(2x)$ & $\!x^{-6}\sin(2x)$ &
	$\!x^{-4}\sin(2x)$ & $\!x^{-2}\sin(2x)$ & $\!\sin(2x)$  \\	\hline
	0 & 0 &
	\, & \, & \, & \, & \, & -1/4 \\
	0 & 2 &
	\, & \, & \, & \, & \, & 7/4  \\
	0 & 4 &
	\, & \, & \, & \, & 105/2 & -21/4  \\
	0 & 6 &
	\, & \, & \, & 10395/4 & -630 & 43/4  \\ \hline
	2 & 2 &
	\, & \, & \, & \, & 9/2 & -13/4  \\
	2 & 4 &
	\, & \, & \, & 315/4 & -120 & 27/4  \\
	2 & 6 &
	\, & \, & 10395/2 & -34965/4 & 945 & -49/4  \\ \hline
	4 & 4 &
	\, & \, & 3675/2 & -13125/4 & 505 & -41/4  \\
	4 & 6 &
	\, & 1091475/8 & -1006425/4 & 202545/4 & -2100 & 63/4 \\
	6 & 6 &
	21611205/2 & -163066365/8 & 19322415/4 & -1210167/4 & 10899/2 & -85/4  \\ \hline
	\end{tabular}
\end{center}
\end{table}
%


\begin{thebibliography}{99}
\bibitem{Effantin}
	J. M. Effantin, J. Rossat-Mignod, P. Burlet, H. Bartholin, S. Kunii and T. Kasuya:
	J. Magn. Magn. Mater. {\bf 47 \& 48}
	(1985) 145.
\bibitem{Erkelens}
	W. A. C. Erkelens, L. P. Regnault, P. Burlet and J. Rossat-Mignod:
	J. Magn. Magn. Mater. {\bf 63\& 64} (1987) 61.
\bibitem{Sakai}
	O. Sakai, R. Shiina, H. Shiba and P. Thalmeier:
	J. Phys. Soc. Jpn. {\bf 66}
	(1997) 3005.
\bibitem{Kusunose}
	H. Kusunose and Y. Kuramoto:
	J. Phys. Soc. Jpn. {\bf 70}
	(2001) 1751.
\bibitem{Kasuya}
	T. Kasuya and D. H. Lyons:
	J. Phys. Soc. Jpn. {\bf 21}
	(1966) 287.
\bibitem{Teitelbaum}
	H. H. Teitelbaum and P. M. Levy:
	Phys. Rev. B {\bf 14}
	(1976) 3058.
\bibitem{Schmidt}
	D. Schmitt and P. M. Levy:
	J. Magn. Magn. Mater. {\bf 49} (1985) 15.
\bibitem{Coqblin}
	B. Coqblin and J. R. Schrieffer:
	Phys. Rev. {\bf 185}
	(1969) 847.
\bibitem{Ohkawa}
	F. J. Ohkawa:
	J. Phys. Soc. Jpn. {\bf 52}
	(1983) 3897.
\bibitem{Shiba}
	H. Shiba, O. Sakai and R. Shiina:
	J. Phys. Soc. Jpn. {\bf 68}
	(1999) 1988.
\bibitem{Takahashi}
	H. Takahashi and T. Kasuya:
	J. Phys. C: Solid State Phys. {\bf 18} (1985) 2755.
\bibitem{Yildirim}
	T. Yildirim, A. B. Harris, A. Aharony and O. Entin-Wohlman:
	Phys. Rev. B {\bf 52}
	(1995) 10239.
\bibitem{IPES}
	Y. Mori, N. Shino, S. Imada, S. Suga, T. Nanba, M. Tomikawa and S. Kunii:
	Physica B {\bf 186-188} (1993) 66.
\bibitem{Chi}
	G. Chiaia, O. Tjernberg, L. Du\`{o}, S. De Rossi, P. Vavassori, I. Lindau, T. Takahashi, S. Kunii, T. Komatsubara, D. Cocco, S. Lizzit, and G. Paolucci:
	Phys. Rev. B {\bf 55} (1997) 9207.
\bibitem{Hirst}
	L. L. Hirst:
	Adv. Phys. {\bf 27} (1978) 231.
\bibitem{Zener}
	C. Zener:
	Phys. Rev. {\bf 82} (1951) 403.
\bibitem{Varshalovich}
	D. A. Varshalovich, A. N. Moskalev, and V. K. Khersonskii:
	{\it Quantum theory of angular momentum}
	(World Scientific, Singapore, 1988).
\bibitem{Brink}
	D. M. Brink and G. R. Satchler:
	{\it Angular momentum}
	(Clarendon press, Oxford, 1968).
\bibitem{Lindgren}
	I. Lindgren J. Morrison:
	{\it Atomic Many-body Theory}
	(Springer-Verlag, Berlin, 1985).
\bibitem{Judd}
	B. R. Judd:
	{\it Operator techniques in atomic spectroscopy}
	(McGRAW-HILL Book Company, New York, 1963).
\bibitem{Zirngiebl}
	E. Zirngiebl, B. Hillebrands, S. Blumenr\"oder, G. G\"untherodt, M. Loewenhaupt, J. M. Carpenter, K. Winzer and Z. Fisk:
	Phys. Rev. B {\bf 30} (1984) 4052.
\bibitem{Khomskii}
	K. I. Kugel and D. I. Khomskii:
	ZhETF Pis. Red. {\bf 15} (1972) 629. [JETP Lett. {\bf 15} (1972) 446].
\bibitem{Shiina}
	R. Shiina, H. Shiba, and P. Thalmeier:
	J. Phys. Soc. Jpn {\bf 66} (1997) 1741.
\bibitem{TandXComment}
	Same result was given in ref.\citen{Shiina} up to rank 3 irreducible tensor operators,
	but we need their projection up to rank 5.
	The irreducible tensor operators $T_Q^P$ and $J_Q^{(P)}$ in ref.\citen{Shiina} differ only by their reduced matrix elements.
\bibitem{SakaiN}
	O. Sakai, R. Shiina, H. Shiba:
	J. Phys. Soc. Jpn {\bf 72} (2003) 1534.
\bibitem{kuramoto-kubo}
	Y. Kuramoto and K. Kubo:
	J. Phys. Soc. Jpn. {\bf 71}
	(2002) 2633.
\bibitem{Goto}
	T. Goto, A. Tamaki, T. Suzuki, S. Kunii, N. Sato, T. Suzuki, H. Kitazawa, T. Fujimura and T. Kasuya:
	J. Magn. Magn. Mater. {\bf 52} (1985) 253;
\bibitem{Nakamura}
	S. Nakamura, T. Goto, S. Kunii, K. Iwashita and A. Tamaki:
	J. Phys. Soc. Jpn. {\bf 63} (1994) 623;
\bibitem{Suzuki}
	O. Suzuki, T. Goto, S. Nakamura, T. Matsumura and S. Kunii:
	J. Phys. Soc. Jpn. {\bf 67} (1998) 4243.
\bibitem{UpperboundL}
	The upperbound of $l$ differs case by case.
	Considering bond density within $j_1 = 5/2$ states, the upperbound of $l$ is 4 because $l$ must satisfy triangle condition, $0 \leq l \leq 2j_1$, and must be an even integer.
\bibitem{Larsen}
	U. Larsen:
	J. Math. Phys. {\bf 21} (1980) 1925.
\bibitem{Ruderman}
	M. A. Ruderman and C. Kittel:
	Phys. Rev. {\bf 96} (1954) 99;
	K. Yosida: {\it ibid.} {\bf 106} (1957) 893;
	T. Kasuya: Progr. Theoret. Phys. (Kyoto) {\bf 16} (1956) 45.
\bibitem{Shiba-Yb4As3}
	H. Shiba, K. Ueda and O. Sakai:
	J. Phys. Soc. Jpn {\bf 69} (2000) 1493.
\end{thebibliography}
\end{document}